\documentclass[]{jfm}

\usepackage{graphicx}
\usepackage{newtxtext}
\usepackage{newtxmath}
\usepackage{natbib}
\usepackage{hyperref}
\hypersetup{
    colorlinks = true,
    urlcolor   = blue,
    citecolor  = black,
}
\usepackage{tabularx}

\usepackage{subfigure}
\usepackage{placeins}
\usepackage{natbib}
\usepackage{color}
\usepackage{rotating}
\renewcommand{\vec}[1]{\mbox{\boldmath$#1$}}

\definecolor{orange}{rgb}{.9,.3,0}

\usepackage{float}
\usepackage{calc}

\title{Reynolds number dependence of Lagrangian dispersion in direct numerical simulations of anisotropic magnetohydrodynamic turbulence}

\author{J. Pratt{\aff{1,}\aff{2}}
  \corresp{\email{jpratt7@gsu.edu}},
  A. Busse\aff{3}
 \and W.-C. M\"uller\aff{4}}

\affiliation{\aff{1}Lawrence Livermore National Laboratory, 4 Ivy Lane, 7000 East Ave, Livermore, CA 94550, USA
\aff{2}Department of Physics and Astronomy, Georgia State University, Atlanta GA 30303, USA
\aff{3}James Watt School of Engineering, University of Glasgow, Glasgow G12 8QQ, United Kingdom
\aff{4}Center for Astronomy and Astrophysics, ER 3-2, TU Berlin, Hardenbergstr. 36, 10623 Berlin, Germany}

\begin{document}
\maketitle

%\date{\today}

\begin{abstract}
Large-scale magnetic fields thread through the electrically conducting
matter of the interplanetary and interstellar medium, stellar interiors, and other astrophysical plasmas, producing anisotropic flows with regions of high-Reynolds-number turbulence. It is common to encounter turbulent flows structured by a magnetic field with a strength approximately equal to the root-mean-square magnetic fluctuations.  In this work, direct numerical simulations of anisotropic
magnetohydrodynamic (MHD) turbulence influenced by such a magnetic field are conducted for a series of cases that have identical resolution, and increasing grid sizes
up to $2048^3$.  The result is a series of closely comparable simulations at Reynolds numbers ranging from 1,400 up to 21,000.
We investigate the influence of the Reynolds number from the Lagrangian
viewpoint by tracking fluid particles and calculating single-particle and two-particle statistics.  The influence of Alfv\'enic fluctuations and the fundamental anisotropy on the MHD turbulence in these statistics is discussed.
Single-particle diffusion curves exhibit mildly superdiffusive behaviors that differ in the direction aligned with the magnetic field and the direction perpendicular to it.
Competing alignment processes affect the dispersion of particle pairs, in particular at the beginning of the inertial subrange of time scales.
Scalings for relative dispersion, which become clearer in the inertial subrange for larger Reynolds number, can be observed that are steeper than indicated by the Richardson prediction.   
\end{abstract}

%\keywords{}
%\keywords{magnetohydrodynamics (MHD) -- turbulence -- magnetic fields -- methods: statistical}

\begin{keywords}
magnetohydrodynamics (MHD) -- turbulence -- magnetic fields -- methods: statistical
%Authors should not enter keywords on the manuscript, as these must be chosen by the author during the online submission process and will then be added during the typesetting process (see \href{https://www.cambridge.org/core/journals/journal-of-fluid-mechanics/information/list-of-keywords}{Keyword PDF} for the full list).  Other classifications will be added at the same time.
\end{keywords}


\section{Introduction}

Only relatively recently have Lagrangian statistics begun to be explored as a tool to understand properties of turbulence in the electrically conducting fluids that are described by magnetohydrodynamics (MHD) \citep{busse_hoho,homann2007lagrangian,busse_astnach,busse2010lagrangian,eyink2013flux,pratt2017extreme,pratt2020intermittency,pratt2020lagrangian}.
In contrast, the study of neutral-fluid turbulence from a Lagrangian viewpoint has a long history \citep[e.g.][]{taylor1922diffusion,richardson1926atmospheric}, with established scaling laws due to Batchelor and Richardson that have been explored both theoretically and experimentally.
In neutral fluids, comprehensive studies have examined how Lagrangian statistics related to dispersion are dependent on the Reynolds number of the flow  \citep{yeung2004relative,yeung2006reynolds,sawford2008reynolds}.  In this work we explore these questions in the more physically complex system of fully nonlinear incompressible MHD turbulence.

Turbulent mixing is investigated through the diffusion and relative
dispersion of fluid tracer particles.  A full description of mixing in a plasma 
depends on the Reynolds number.
Relative dispersion in neutral fluids quantifies the
mixing of smoke and pollutants in the atmosphere, or of micro-plastics and debris
in the oceans.  In electrically conducting fluids, diffusion, dispersion, and the mixing processes they represent determine how
fusion products from the core of a star are mixed into the star's outer-layers, thereby changing the course of stellar evolution.  Relative dispersion
also represents the spreading and mixing of plasma in the interstellar or interplanetary medium, and affects how energetic particles and cosmic rays are transported.  In these contexts, Reynolds numbers are predicted that are many orders higher than can be achieved presently by direct numerical simulations (DNS) that fully resolve the turbulent motions.  However, the features of diffusion and dispersion captured by fluid particles in DNS of MHD turbulence provide information relevant to turbulent mixing in these astrophysical applications  \citep[e.g.][]{zahn1993mixing,heyer2004universality,utomo2019origin}.  The details of how these statistics change as the Reynolds number increases allow us to make predictions for realistic mixing.

In astrophysical settings, the magnitude of the large-scale magnetic field is often moderate.  
For example, in the solar wind, ion foreshock, and magnetosheath, ranges have been reported such that the magnetic field is between one and 2.5 times the RMS fluctuations, i.e. 
$ 1 \leq B_0 / B_{\mathsf{RMS}}  \leq 2.5 $ (see table 1 of \citet{zimbardo2010magnetic}).
In this work, we examine a system with a weak anisotropy caused by such a large-scale magnetic field, and select the situation where the magnetic field is equal to the average RMS magnetic field fluctuations, i.e. $B_{\mathsf{RMS}} \approx B_0$. 

This work is structured as follows.   In Section \ref{secsim} we describe in detail the simulations performed.  In Section \ref{secrescrit} we introduce a new resolution criterion for anisotropic MHD turbulence simulations, a necessity for producing a set of direct numerical simulations that can be closely compared.   In Section \ref{usualsect} we present diffusion curves, a single-particle Lagrangian statistic.  In Section \ref{secresults2} we present several statistics derived from pairs of tracer particles, focused on understanding the relative dispersion.  In Section \ref{secdiscussion} we summarize our findings and draw broader conclusions from the statistics presented.

\section{Simulations \label{secsim}}

We investigate the effect of the Reynolds number on statistically stationary,
forced, homogeneous, incompressible magnetohydrodynamic turbulence in the presence of a moderate static magnetic field. 
 This magnetic field has a constant value and direction throughout the simulation.  
Locally the magnetic field also experiences time-dependent fluctuations which possess zero mean.  The strength of these fluctuations can be measured with respect to the strength of the imposed magnetic field.  A global average of the magnetic field would return the value due to the imposed field, and therefore the term \emph{mean magnetic field} best describes the type of field imposed.  Such a magnetic field has alternatively been referred to an external magnetic field, or a guide field, although those terms are less specific to the way that the field has been imposed.

In each direct numerical simulation, we solve the non-dimensional equations for magnetohydrodynamics: 
\begin{eqnarray} \label{realbmhdc}
\frac{\partial \vec{\omega} }{\partial t} &-& \nabla \times (\vec{\mathrm{v}} \times \vec{\omega} +  \vec{j} \times \vec{B})
=  \hat{\nu} \nabla^2 \vec{\omega}  + \vec{f}^{\omega} ~,
\\ \label{realbmhdb}
\frac{\partial \vec{B} }{\partial t} &-& \nabla \times (\vec{\mathrm{v}} \times \vec{B}) =  \hat{\eta} \nabla^2 \vec{B}+ \vec{f}^{b}~,
\\ \label{realbmhdinc}
 &~&\nabla \cdot \vec{B} = 0~~~~,~~~~~\nabla \cdot \vec{\mathrm{v}} = 0~,
\end{eqnarray}
using a pseudospectral method in a simulation volume with
periodic boundary conditions. These equations include terms for the solenoidal
velocity field $\vec{\mathrm{v}}$, vorticity $\vec{\omega}=\nabla\times\vec{\mathrm{v}}$,
magnetic field $\vec{B}$, and current $\vec{j}=\nabla\times\vec{B}$. 
Each of the quantities in eqs.~\eqref{realbmhdc} -- \eqref{realbmhdinc} has been non-dimensionalized using relevant time and length scales, commonly referred to as Alfv\'enic units. Two dimensionless
parameters, $\hat{\nu}$ and $\hat{\eta}$, appear in the equations. They
derive from the kinematic viscosity $\nu$ and the magnetic diffusivity
$\eta$.  A fixed time-step and a low-storage third-order Runge--Kutta
method \citep{will80} are used for the time-integration. The mean magnetic field is designated by $B_0$, and points purely in the positive $z$-direction.  
At any point in time it has a value close to unity with respect to the RMS of the magnetic field fluctuations; a long-time average produces $B_0  \approx B_{\mathsf{RMS}}$. 
The Alfv\'en ratio is approximately unity for all
simulations discussed in this work; we calculate this ratio as $r_{\mathrm{A}} = \langle
\mathsf{E_v} /\mathsf{E_b} \rangle$ from the kinetic energy per unit mass
$\mathsf{E_v}= \vec{\mathrm{v}}^2/ 2$ and the energy per unit mass contained in the magnetic fluctuations
$\mathsf{E_b}= B_{\mathsf{RMS}}^2 / 2$. In the Alfv\'en ratio, the brackets indicate an average over time; this is the simulation time during which the dynamics of Lagrangian tracer particles are examined.  The implications of an Alfv\'en ratio of one are that our simulations
include the full nonlinear interaction of velocity and magnetic fields that contribute with equal weight.

To maintain the turbulence in a statistically stationary steady state, the vorticity and magnetic
fields are forced on the largest scales of the simulation volume using a
deterministic forcing, applied in Fourier space, that also allows the largest scale motions of the
system to evolve.  Deterministic forcing has the advantage that no stochastic source of fluctuations is introduced on large scales.
  We call the deterministic forcing method that we use \emph{homogeneous
forcing}; it is distinct from forcing methods used in several earlier works on Lagrangian MHD turbulence \citep{busse_hoho,homann2007lagrangian,busse_astnach,busse2010lagrangian}
 but identical to the forcing method used in \citet{pratt2020lagrangian}.
 Homogeneous
forcing establishes a constant injection of energy at large
scales.  In eqs.~\eqref{realbmhdc} and \eqref{realbmhdb} forcing terms
$\vec{f}^{\omega}$ and $\vec{f}^{b}$ are introduced which
are non-zero only for the wave-vector shells $1 \leq |\vec{k}| \leq
2.5$.  The forcing terms are defined
\begin{eqnarray}
\vec{\hat{f}}^{\omega}(\vec{k},t)
& = &\gamma_{f,\omega}
\frac{\hat{\vec{\omega}}(\vec{k},t)}{|\hat{\vec{\omega}}(\vec{k},t)|^2} ~,
\\
\vec{\hat{f}}^{b}(\vec{k},t)& = &\gamma_{f,b}
\frac{\hat{B}(\vec{k},t)}{|\hat{B}(\vec{k},t)|^2}~.
\end{eqnarray}
Variables with hats are used to describe the forcing because it is applied in Fourier space.  The constants $\gamma_{f, \omega}$ and
$\gamma_{f, b}$ regulate the energy injection.    These two
forcing constants are set equal in our simulations, and are identical across all of the simulations presented in this work.  
 By forcing both fields equally, we achieve approximate equipartition of kinetic and magnetic energy at all scales,
and study the regime of MHD turbulence defined by the full non-linear interaction of the velocity and magnetic fields.
This is in contrast to studies that force only the velocity field, oriented toward understanding the dynamo \citep[e.g.][]{brandenburg2018varying,mckay2017comparison}.
In natural settings such as molecular clouds \citep[][]{hennebelle2012turbulent,heiles2005millennium}, and the solar wind
\citep[][]{boldyrev2012residual,muller2005spectral} kinetic and magnetic energies are commonly observed to be close to equipartition, so 
this choice is a physically realistic one.

To inhibit the emergence of states dominated by Els\"{a}sser  \citep{elsasser1950hydromagnetic} positive ($z^+$) or negative ($z^-$) interactions, the cross helicity of the forced modes is set to zero.   This is accomplished as part of the forcing scheme by enforcing orthogonality between the magnetic field vector and the velocity field vector \citep[see e.g.,][]{muller2012inverse}.   
The magnitude of the total cross helicity, normalized by $\sqrt{\mathsf{E_v}}\sqrt{\mathsf{E_b}}$ never rises above $0.095$ in the simulations examined in this work.
 This prevents  the MHD turbulent system from becoming imbalanced, which can lead to a break-down of the non-linear energy
cascade \citep[as discussed in][]{biskampbook}. 
In addition, the level of total magnetic helicity, normalized by $\mathsf{E_b}^2/k_{\mathsf{forcing}}$, never rises above $5 \cdot 10^{-4}$. 
 For a system in a quasi-stationary state, homogeneous forcing is expected
to disturb the natural turbulent flow only mildly; this forcing method has been examined in \citet{ghosal1995dynamic,vorobev2005anisotropy}.  Based on the forcing, our simulations are consistent with strong turbulence as described by \citet{perez2007weak} and \citet{verdini2012transition}.  Large-scale Alfv\'en waves are permitted and are
observed when homogeneous forcing is used.

\subsection{Lagrangian tracer particles}

The positions of Lagrangian tracer particles are initialized in a homogeneous random distribution at a time when the turbulent flow has attained
a statistically stationary state.   The approximate number of tracer particles deployed in each simulation, in millions, is given in table~\ref{simsuma}.  The particle numbers we use are comparable to earlier MHD turbulence studies \citep{busse_hoho,homann2007lagrangian,busse_astnach,pratt2020lagrangian}, and are larger than those used in \citet{yeung2006reynolds} and \citet{sawford2008reynolds} to study homogeneous isotropic hydrodynamic turbulence. The consequence is that the Lagrangian statistics that we produce in this work are well-resolved in space.   At each time-step
the particle velocities are interpolated from the instantaneous Eulerian
velocity field using a tricubic polynomial interpolation scheme
\citep{lekien2005tricubic,homann2007impact}.  Particle positions are calculated by
numerical integration of the equation of motion using a low-storage
third-order Runge--Kutta method that matches the one used for the Eulerian field integration.   We record particle information every four time steps, at approximate time intervals of $0.1 \tau_{\eta}$, so that the Lagrangian statistics produced are also well-resolved in time.  The hydrodynamic simulation that we perform for comparison, simulation  3H, uses a slightly larger time interval of $0.2 \tau_{\eta}$.  
Particles are initially arranged in tetrads
with a reference particle at the vertex and three other
particles aligned along each Cartesian direction at fixed separations
from the reference particle.  
Each dispersion result is defined by this initial separation distance.
The smallest initial separation distances that we deploy are well-resolved by the grids used in each simulation; this will be quantified in the section on the resolution.  
Each simulation is run for at least 400 of the Kolmogorov time scale $\tau_{\eta}$. 
At large times, the particles that make up a pair can reach a separation that is on the scale of the simulation volume.   
For pairs of particles that have separated that far, the probability is low for a statistically significant number to reapproach each other again.
Although the forcing is applied on length scales comparable to the simulation volume, the velocity field experienced by pairs of Lagrangian tracer particles separated by the size of the simulation volume is generally decorrelated, and the two particles move approximately independently from each other.

\section{Resolution of anisotropic MHD turbulence \label{secrescrit}}

\begin{figure}
\resizebox{2.8in}{!}{\includegraphics{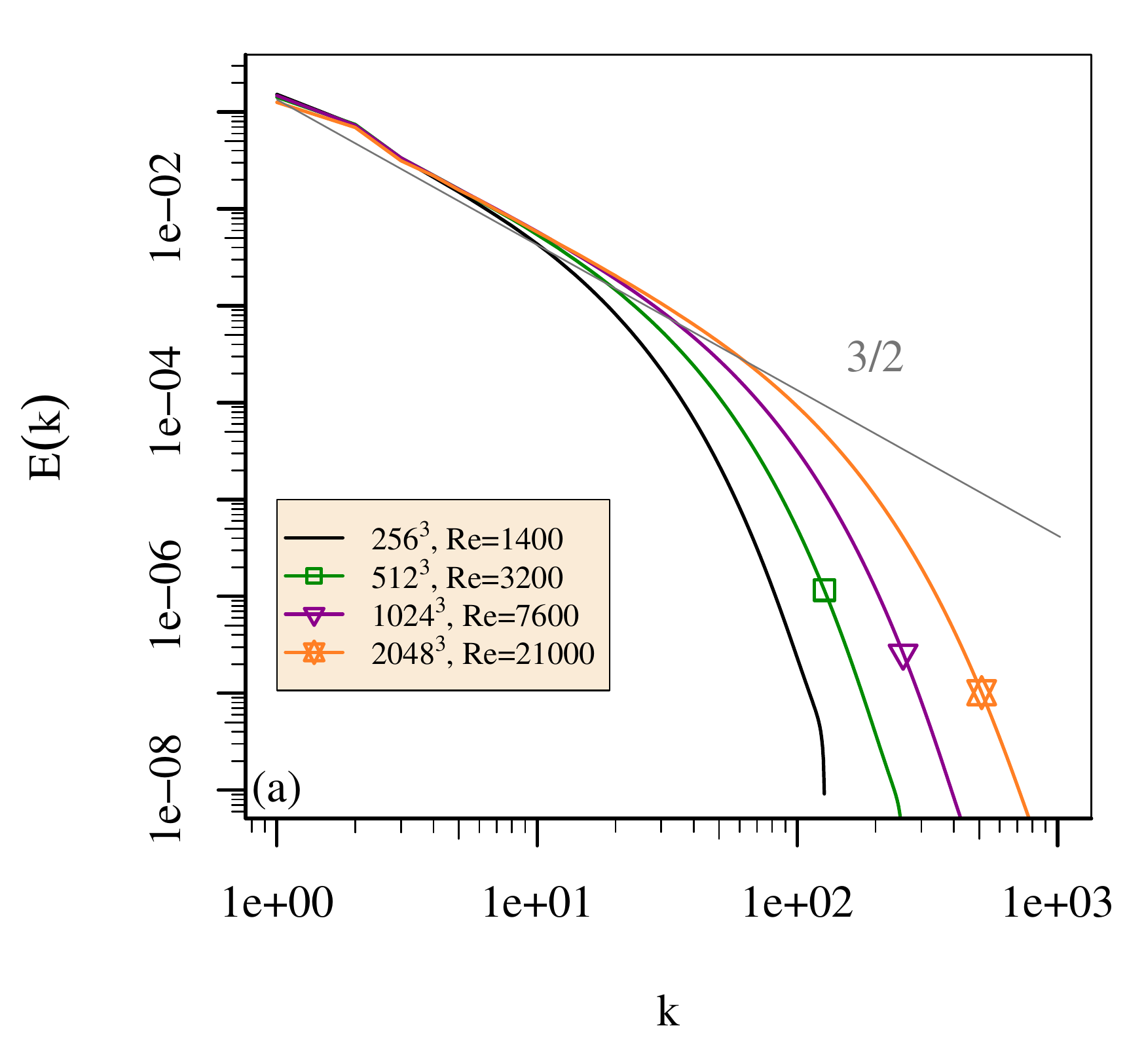}}
\resizebox{2.8in}{!}{\includegraphics{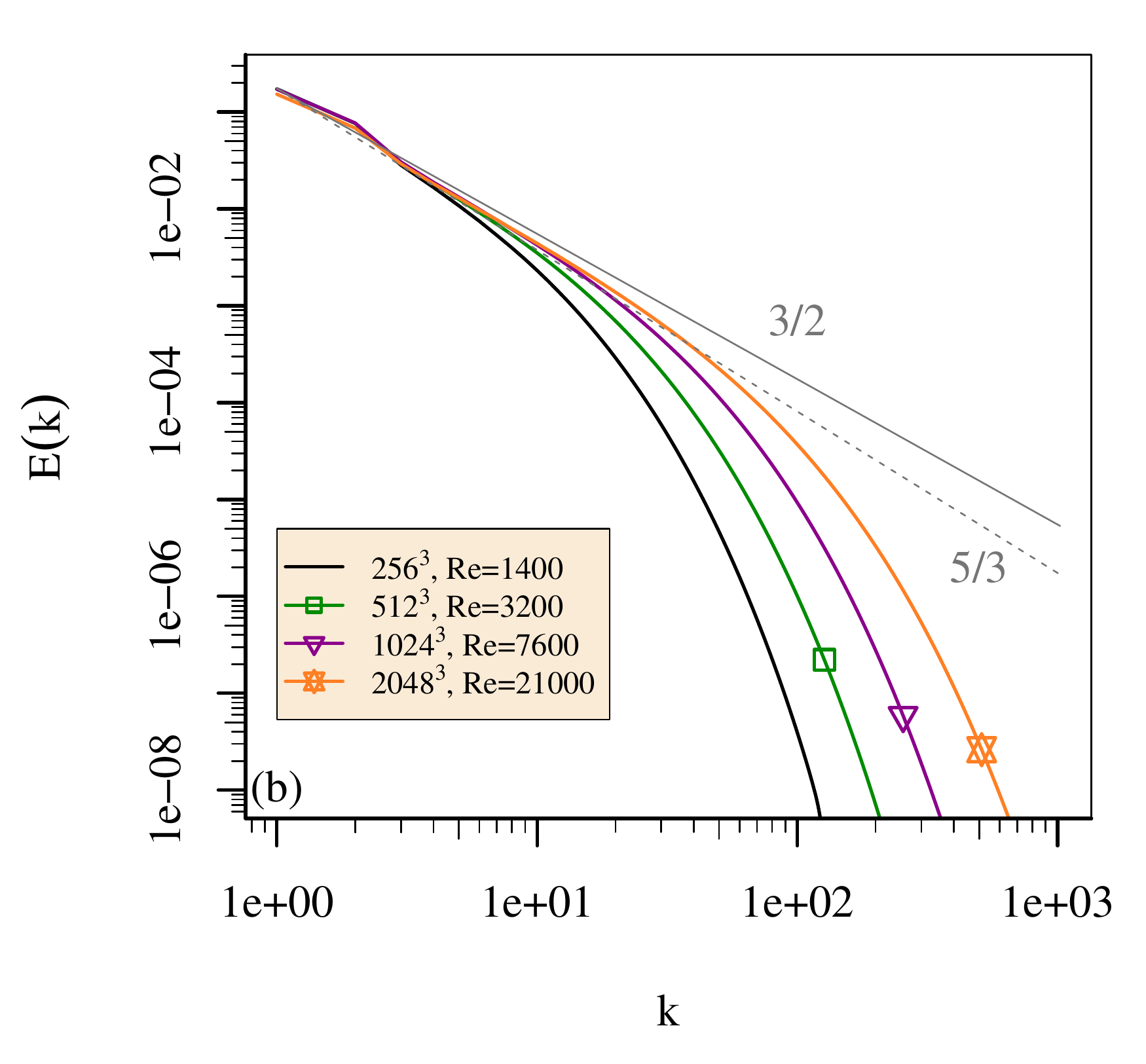}}
\caption{Time-averaged kinetic energy spectra for the four MHD simulations described in
table~\ref{simsuma}. The spectra are calculated for a one-dimensional wavevector $k$ taken in the (a) $x$-direction (perpendicular to the mean magnetic field), and (b) $z$-direction (parallel to the mean magnetic field).  
Grey lines indicate theoretical scaling laws relevant to the inertial range for MHD turbulence.
\label{figeulerian}
}
\end{figure}

In anisotropic MHD turbulence, attention must be
given both to the shape of the simulation volume and the ability
to resolve the smallest relevant scales of the turbulent dynamics.  At both large scales and small scales, differences exist between the energy spectra in the directions parallel and perpendicular to the
mean magnetic field (see figure~\ref{figeulerian}).
The asymptotic spectral scaling exponents of the one-dimensional energy spectra are approximately -1.5 (perpendicular) and -1.6 (parallel). 
These spectra are defined as $E(k)=\int d^3\vec{k}' E(\vec{k}') \delta(|\vec{k}'\cdot \hat{\vec{e}}_k|-k)$, where
the wavenumber $k$ runs along the direction given by the appropriately chosen and fixed unit vector $\hat{\vec{e}}_k$.
These scalings are consistent with the current understanding of inertial-range dynamics of incompressible MHD turbulence, which is based on
the concepts of critical balance of characteristic timescales parallel and perpendicular to the local magnetic field 
\citep{goldreich_sridhar2,malletrefinedcb} and dynamical
alignment of fluctuations of velocity and magnetic field \citep{boldymodelII}, in combination with specific Log-Poisson intermittency corrections
\citep{chandranikintermitt,malletstrongMHDmod}. We provide these scalings purely for the sake of completeness; they should be viewed with care. The main caveat is the limited width of the inertial scaling range, 
 a consequence of the well-resolved dissipation region required for Lagrangian small-scale statistics. Another relevant issue is 
that anisotropic MHD energy transfer for systems with moderate mean magnetic fields should be analyzed in a local frame of reference aligned with the local
mean magnetic field, which includes contributions from magnetic fluctuations, instead of the global mean field $B_0$.  Such an analysis of spectral transfer is outside the scope of this Lagrangian investigation.
For the present work, the following simple observations of MHD spectral dynamics seem sufficient to characterize the system investigated.
As the energy cascade proceeds to smaller scales, the eddies
become more elongated in the 
direction parallel to the magnetic field; the amount of elongation is dependent on the length-scale of motions in that direction, which in turn is dependent on the strength of the magnetic field \citep[e.g. as discussed in][]{cho2000anisotropy,schekochihin2008mhd,verdini2015anisotropy}.  We explored the corresponding time-scale dependence of this anisotropy from the Lagrangian point of view in \citet{pratt2020lagrangian}.
Due to the elongation of eddies in the direction aligned with the magnetic field, the smallest length scales relevant for the study
of turbulence are different from those in the perpendicular direction.   At the largest scales, correlation lengths are also longer in the
direction aligned with the mean magnetic field than in the perpendicular direction.

\subsection{Resolution of the largest scales of anisotropic MHD turbulence}

Corresponding to these correlation lengths, velocity structures have different characteristic scales in 
the directions parallel and perpendicular to the mean magnetic field.
To assure that these velocity structures are both resolved, 
we consider how the periodic simulation volume should be shaped.
 We solve the magnetohydrodynamic eqs.~\eqref{realbmhdc} -- \eqref{realbmhdinc}
  in a simulation volume with sides of length $2 \pi$ in the $x$ and
$y$ directions, that are perpendicular to the mean magnetic field. Using a simulation volume that is elongated in the $z$-direction
 is common for simulations of anisotropic MHD that examine a strong mean magnetic field.
 To determine the necessary length in
the $z$-direction, we consider the correlation length of the
velocity field in each direction.  
   In test simulations with a stronger mean
magnetic field, we measure a correlation
length of the velocity field $L_{\mathsf{c},\parallel}$ in the direction parallel to $B_0$ that is significantly larger than in the perpendicular direction.  This measurement agrees with previous
results
\citep[e.g.][]{chandran2008strong,boldyrev2005spectrum,
cho2002simulations}.  For the strong turbulence regime, a relative
change in length scales can be predicted from the premise of critical
balance \citep{goldreich1995toward}.  Using this premise, the ratio of
the largest-scale wavenumbers ($k \sim 2 \pi / L$) in the perpendicular and
parallel directions grows linearly with the magnitude of the mean
magnetic field $B_0$:
\begin{eqnarray}\label{criticalbalance}
k_{\parallel} B_0 \sim k_{\perp} B_{\mathsf{RMS}} ~.
\end{eqnarray}
Thus for the largest scales of the flow, critical balance predicts that the ratio of parallel and perpendicular length scales should grow linearly with $B_0$.
 The critical balance relation in eq.~\eqref{criticalbalance} 
implies that the non-linear eddy-turnover time is of the order of 
the Alfv\'{e}n time $\tau_{\mathsf{NL}} \sim \tau_{\mathsf{A}}$.    For the $B_0=1$ simulations examined in this work, 
  the box length in the parallel direction is much larger than the
 parallel correlation length, $L_z \gtrsim 10 L_{\mathsf{c},\parallel}$.  We therefore
 are able to use a cubic simulation volume with no elongation for all simulations in table~\ref{simsuma}.

\subsection{Resolution of the smallest scales of anisotropic MHD turbulence}

We revisit the resolution criterion commonly
used to measure whether isotropic turbulence is sufficiently resolved, and we 
extend this criterion to the anisotropic case.   The smallest length and time scales that characterize
Navier--Stokes turbulence are defined in terms of the kinetic energy dissipation rate $\epsilon_{\mathrm{v}}$ and the kinematic viscosity.
These are the Kolmogorov microscale
$\mathsf{\eta_{kol}=(\nu^3/\mathsf{\epsilon_{\mathrm{v}}})^{1/4}}
$ and the Kolmogorov time-scale
$\mathsf{\tau_{\eta}=(\nu/\mathsf{\epsilon_{\mathrm{v}}})^{1/2}}$.
To test whether a direct numerical simulation adequately resolves the smallest
physically relevant spatial scales of turbulence, the Kolmogorov microscale $\eta_{\mathsf{kol}}$ is typically multiplied by 
the highest wavenumber resolved in a simulation $k_{\mathrm{max}}$.
  This provides a nondimensional number
that indicates how well the grid spacing resolves this smallest physical length scale.   The ``standard'' criterion for adequate spatial resolution for a simulation in the case of
homogeneous isotropic turbulence  \citep{pope2000turbulent,yeung1989lagrangian,donzis2008dissipation,yeung2018effects} is then
\begin{eqnarray}\label{pope-criterion}
k_{\mathrm{max}} \mathsf{\eta_{kol}} \gtrapprox 1.5~.
\end{eqnarray}
The presence of a mean magnetic field
makes the gradients of the turbulent fields higher in the perpendicular
directions compared to the gradients in the parallel direction.
Therefore, we need to extend the criterion in eq.~\eqref{pope-criterion} to
ensure that the smallest eddies are well-resolved both for the mean field parallel and perpendicular direction.

To extend this resolution criterion, we examine the definition of the kinetic energy dissipation rate, which is related to gradients in the velocity field
\begin{eqnarray}
\epsilon_{\mathsf{v}}=
\frac{1}{2}\sum_{i,j}\nu \left\langle \left(\frac{\partial \mathrm{v}_i}{\partial x_j}\right)^2\right\rangle ~.
\end{eqnarray}
In Fourier space this is expressed
\begin{eqnarray}\label{eps-fourier}
\epsilon_{\mathsf{v}}=
\sum_{\vec{k}} \nu |\vec{k}|^2 \sum_{i=1}^{3}\langle \hat{\mathrm{v}}_i^{\ast}(\vec{k},t) \hat{\mathrm{v}}_i(\vec{k},t)\rangle ~.
\end{eqnarray}
By separating the modulus of the wave-vector in eq.~\eqref{eps-fourier} into
components parallel $k_{\parallel}=\vec{k}\cdot (B_0 \vec{\hat{z}})/B_0$
and perpendicular $\vec{k}_{\perp}=\vec{k}\times (B_0 \vec{\hat{z}})/B_0$
to the direction of anisotropy, the kinetic energy dissipation rate can
be split into contributions that arise from gradients parallel and
perpendicular to the mean magnetic field.  We define these components
\begin{eqnarray}
\epsilon_{\mathsf{v}} &=&\frac{2}{3}\epsilon_{\mathsf{v},\perp} + \frac{1}{3}\epsilon_{\mathsf{v},\parallel}~,
\\
\epsilon_{\mathsf{v},\perp}&=&\frac{3}{2} \sum_{\vec{k}} \nu |\vec{k}_{\perp}|^2 \sum_{i=1}^{3}\langle \hat{\mathrm{v}}_i^{\ast}(\vec{k},t) \hat{\mathrm{v}}_i(\vec{k},t)\rangle~,
\\
\epsilon_{\mathsf{v},\parallel} &=& 3 \sum_{\vec{k}} \nu k_{\parallel}^2 \sum_{i=1}^{3}\langle \hat{\mathrm{v}}_i^{\ast}(\vec{k},t) \hat{\mathrm{v}}_i(\vec{k},t)\rangle~.
\end{eqnarray}
In the case of homogeneous isotropic turbulence, these definitions recover
$\epsilon_{\mathsf{v},\perp}=\epsilon_{\mathsf{v},\parallel}=
\epsilon_{\mathsf{v}}$. In the case of a finite mean magnetic field,
gradients perpendicular to the mean magnetic field will be larger than
gradients parallel to the mean magnetic field, leading to
$\epsilon_{\mathsf{v},\perp}> \epsilon_{\mathsf{v}}$ and
$\epsilon_{\mathsf{v},\parallel}< \epsilon_{\mathsf{v}}$. 

Two different length scales, for the direction perpendicular and
parallel to the mean magnetic field, can be defined from $\epsilon_{\mathsf{v},\perp}$ and $\epsilon_{\mathsf{v},\parallel}$.  These are
\begin{eqnarray}
 \label{defkolmicroperp}
\eta_{\mathsf{kol},\perp}&=&\left(\hat{\nu}^3/\epsilon_{\mathsf{v},\perp} \right)^{1/4}~,
\\ \label{defkolmicropar}
\eta_{\mathsf{kol},\parallel}&=&\left(\hat{\nu}^3/\epsilon_{\mathsf{v},\parallel} \right)^{1/4} ~.
\end{eqnarray}
In the isotropic case, these length scales are equal to the Kolmogorov
length scale. Using $\eta_{\mathsf{kol},\perp}$ and
$\eta_{\mathsf{kol},\parallel}$ a generalized version of the small-scale resolution
criterion in eq.~\eqref{pope-criterion} 
can be defined for homogeneous anisotropic turbulence.  This combined criterion for adequate
resolution of the smallest scales both parallel and perpendicular to the
mean magnetic field direction is
\begin{eqnarray}\label{rescritperp}
k_{\mathsf{max},\perp} \eta_{\mathsf{kol},\perp} &\gtrapprox&1.5 ~,
\\ \label{rescritpar}
k_{\mathsf{max},\parallel} \eta_{\mathsf{kol},\parallel} &\gtrapprox&1.5  ~.
\end{eqnarray}
These two criteria are fulfilled in all simulations discussed in this work, and their values
are listed in table~\ref{simsuma}.   When we use these criteria to set-up our simulations, the smallest scales
of anisotropic MHD turbulence appear to be well-resolved in both directions.  
For example, \citet{perez2014scaling} notes that the numerical effects of insufficient small-scale resolution are a steepening of the spectrum at intermediate scales and a flattening
closer to the grid scale.  Those effects are not observed in the spectra of our simulations in figure~\ref{figeulerian}.
The smallest of initial separation distances for pairs of Lagrangian tracer particles are set to $2 \eta_{\mathsf{kol},\perp}$.  Because the Kolmogorov length scale in this direction
is well-resolved, so are these particle separations.

\subsection{Reynolds number for anisotropic MHD turbulence}

For a homogeneous isotropic system, the Reynolds number is standardly defined from the kinetic energy, the viscosity, and a characteristic length
scale 
\begin{eqnarray}\label{defisore}
\mathsf{Re= \langle E_v^{1/2} \mathsf{L_E}\rangle /\nu}~.
\end{eqnarray}
The characteristic length scale $\mathsf{L_E}$ is defined as a
dimensional estimate of the size of the largest eddies,
$\mathsf{L_E}=\mathsf{E_v}^{3/2}/\mathsf{\epsilon_{\mathrm{v}}}$.  This
length scale is summarized in table~\ref{simsuma} for each of our simulations.

To calculate the Reynolds number for anisotropic flows, we use the more general definition of the Reynolds number  \citep[see Chapter 6.1.2 of ][]{pope2000turbulent}
\begin{eqnarray}\label{defanisore}
\mathsf{Re}= c~ (\eta_{\mathsf{kol,\perp}} / \mathsf{L_F})^{-4/3}~,
\end{eqnarray}
where $\mathsf{L_F}$ is a forcing length scale, and $c$ is a constant that must be determined.
Our method of homogeneous forcing affects a minimum length scale
\begin{eqnarray}\label{defforcinglength}
 \mathsf{L_F}=2 \pi/ k_{\mathsf{F,max}} = 0.8 \pi ~.
\end{eqnarray}
We determine the constant $c$ by comparing the definitions of the
Reynolds number  in eq.~\eqref{defisore} with that in eq.~\eqref{defforcinglength} for an isotropic $B_0=0$ simulation;
this produces a value of the constant $c \approx 1.16$.  The Reynolds number
calculated using eq.~\eqref{defanisore} is summarized in table~\ref{simsuma} for each of our simulations.  
For an isotropic flow, the standard Reynolds number based on the Kolmogorov microscale, $\mathsf{Re}$, and the Taylor-scale Reynolds number, $\mathsf{Re}_{\lambda}$, provide equivalent information \citep[see for example ][]{pope2000turbulent,sawford2008reynolds}, and are related by 
\begin{eqnarray}
15~ \mathsf{Re} = \mathsf{Re}_{\lambda}^2~.
\end{eqnarray}
Thus the Reynolds numbers in table~\ref{simsuma} can be translated to Taylor-scale Reynolds numbers; for simulation 4, we find $\mathsf{Re}_{\lambda} \approx 560$. 
The magnetic Reynolds number is defined from the
Reynolds number and the magnetic Prandtl number, i.e.
$\mathsf{Re_m}=\mathsf{Pr_m Re}$.  In all simulations in this work, the
magnetic Prandtl number $\mathsf{Pr_m}=1$ so that the magnetic Reynolds
number is equal to the Reynolds number.  

Table~\ref{simsuma} provides an overview of the Reynolds number and other fundamental parameters.   Among the fundamental parameters in this table are also two time scales: the Kolmogorov time scale $\tau_{\eta}$, which is associated with the smallest scales of motion, and a time scale associated with the largest scales of motion, called the large-eddy turnover time $\mathsf{T_E} =\mathsf{E_v}/\mathsf{\epsilon_{\mathrm{v}}}$.
 Our measurements for the ratio of these time scales, $\tau_{\eta}/\mathsf{T_E}$ decrease as $\mathsf{Re}^{-1/2}$, as expected for isotropic hydrodynamic turbulence \citep[see Chapter 6.1.2 of ][]{pope2000turbulent}.

\begin{table*}
\begin{center}
\def~{\hphantom{0}}
\begin{tabular}{lcc|ccccccc}
sim. no.                 & grid size $N^3$ & $N_p$ (millions)  & $B_{\mathsf{RMS}} $  & $\tau_{\eta}$($10^{-2}$) & $\mathsf{L_E}$   & $\mathsf{T_E}$
%                            &  $\eta_{\mathsf{kol}}$ ($10^{-3}$) &  $\eta_{\mathsf{kol,||}}$ ($10^{-3}$) &  $\eta_{\mathsf{kol,\perp}}$ ($10^{-3}$)     
                            &  $k_{\mathsf{max},\perp} \eta_{\mathsf{kol},\perp}$ & $k_{\mathsf{max},\parallel} \eta_{\mathsf{kol},\parallel} $
                            & $\mathsf{Re}$ 
%                 grid size & Np      & B      % tau_eta %LE    %TE  % kmaxeta % % kmaxeta % Re
\\ \hline
1                  & $256^3$  & $1$ & 0.90 & 12.59  &  3.32   &   5.00     &  1.61   & 1.99  &  1400 %TNOVIS1
\\ \hline
2                 & $512^3$ &  $8$ &0.98 & 8.21  &  3.49      &    5.16     &   1.70  & 2.11  &  3200  % KNOVIS1 3180
 \\ \hline
3                  & $1024^3$ & $8$ & 1.03 & 5.28  &  3.67    &    5.32      &  1.77   & 2.18  &  7600  % ANOVIS1
 \\ \hline
3H                & $1024^3$ & $8$ & 0.0 & 4.78  &   3.56     &    4.88     &  1.77   & 1.77  &  10,400  % HNOVIS1
 \\ \hline
4                  & $2048^3$  & $8$ & 1.07 & 3.17  &  3.35   &    5.00      &  1.64   & 2.04   &  21,000  %SNOVIS1 21,200
\\ \hline
\end{tabular}
\caption{Simulation parameters:
the simulation number, the Eulerian grid size $N^3$, and the number of tracer particles $N_p$ is provided for each simulation.  The Kolmogorov time scale $\tau_{\eta}$, large-eddy length scale $\mathsf{L_E}$, the large-eddy turnover time $\mathsf{T_E}$, and the time-averaged root-mean-square of magnetic fluctuations $B_{\mathsf{RMS}}$ are given.  The magnitude of the mean magnetic field is $B_0=1$ for case 1,2,3,4.   
 The resolution in the perpendicular direction $k_{\mathsf{max},\perp} \eta_{\mathsf{kol},\perp}$ and in the parallel direction $k_{\mathsf{max},\parallel} \eta_{\mathsf{kol},\parallel}$ are provided.   The Reynolds number is calculated as described in eq.~\eqref{defanisore} using the perpendicular Kolmogorov microscale $\eta_{\mathsf{kol,\perp}}$.
   All simulations take place in a cubic simulation volume, and flow statistics are gathered for at least $400 \tau_{\eta}$.    The simulation 3H is a purely hydrodynamic simulation performed for comparison with simulation 3. \label{simsuma}
 }
\end{center}
\end{table*}

\section{Single-particle Lagrangian Diffusion \label{usualsect}}

The examination of single-particle diffusion curves, the average square distance that a particle has moved from its initial position, is common in Lagrangian studies of turbulence.  We designate this quantity by $\langle \Delta^2(t) \rangle$, where the brackets indicate an average over all Lagrangian tracer particles.
In anisotropic MHD turbulence, such diffusion curves were first studied by \citet{busse_astnach}, who examined the influence of mean magnetic fields of magnitude $B_0  = 2$ and 5 in units of $B_{\mathsf{RMS}}$.  Our present results are physically distinct from that work in three ways: (1) we examine a significantly weaker mean magnetic field, (2) we use a different forcing method, and (3) our simulations extend to a higher Reynolds number.
 \begin{figure}
\resizebox{2.8in}{!}{\includegraphics{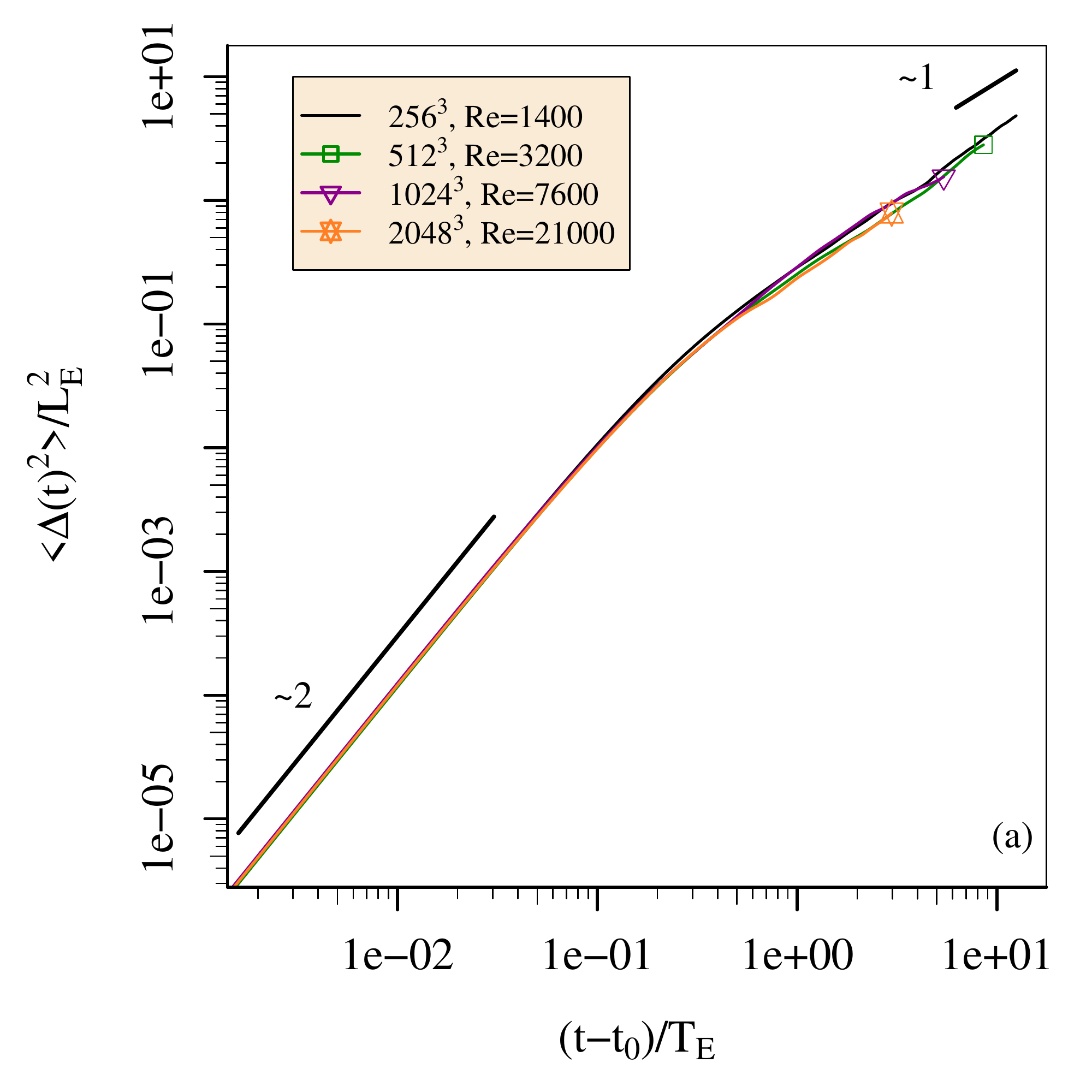}}
\resizebox{2.8in}{!}{\includegraphics{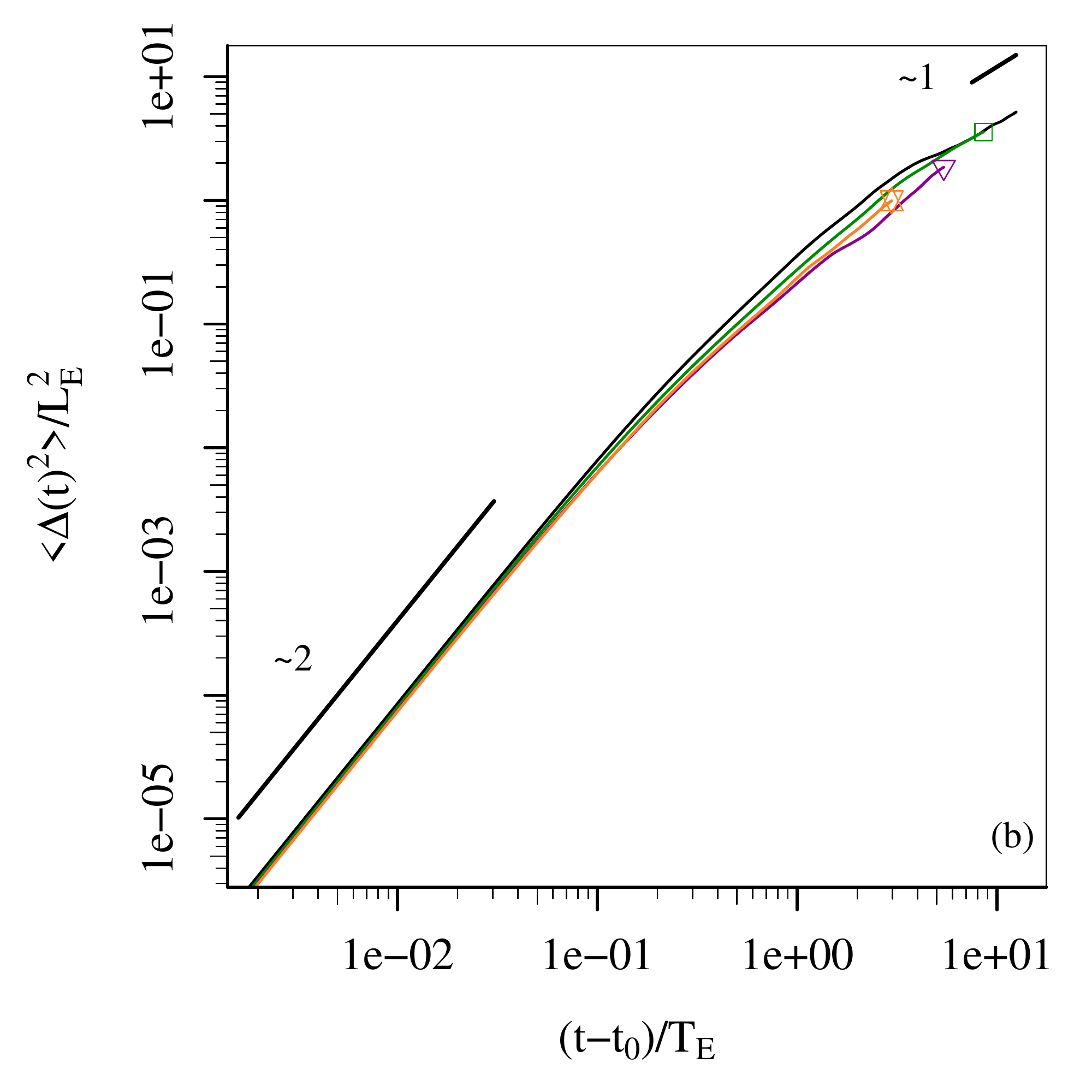}}
\caption{ Average square separation of particles from their
initial position $\langle \Delta^2(t) \rangle$, for (a) separations perpendicular to the mean magnetic
field and (b) separations aligned with the mean magnetic field.  Each curve represents an average over at least three independent initial times.   Distance is measured in units of the large-eddy length scale $\mathsf{L_E}$ and time is measured in units of the large-eddy turnover time $\mathsf{T_E}$.  The straight black lines indicate the scaling laws that are theoretically predicted, with the scaling exponent labeled.
\label{figspdiffusion}
}
\end{figure}

If diffusion curves are calculated from a single initial time, they can be influenced by idiosyncrasies of the flow at that time.  We produce our diffusion curves by averaging the results for several independent initial times using the conventional averaging methods described in \citet{dubbeldam2009new}, so that they are not dependent on any single initial state of the flow.   Diffusion, as a single particle statistic, is also notoriously sensitive to large-scale flow features because each particle's separation is measured relative to a fixed point in space.  
Large-scale flows are able to sweep along large numbers of Lagrangian particles, affecting the outcome of diffusion curves.  Because of this large-scale sweeping, diffusion statistics cannot be formulated to provide information about a specific length scale and the related time scale. 

Since diffusion is dominated by the largest-scale fluctuations of the system that carry most of the kinetic energy,
the natural time-scale to normalize the time is the large-eddy turnover time $\mathsf{T_E}$.
For a fixed integral scale of turbulence, the Kolmogorov time decreases as the extension of the inertial range increases.
In a statistically isotropic Navier-Stokes setting, dimensional analysis relates this large-scale quantity to the Kolmogorov time scale  
$\mathsf{T_E} \sim \tau_{\eta} \mathsf{Re}^{1/2}$ \citep{ishihara_gotoh_kaneda,pope2000turbulent}.  
 This relation is responsible for the later arrival of particles in the diffusive regime with increasing Reynolds number reported in \citet{sawford1991reynolds}.
In the MHD case, we reproduce this later arrival of particles in the diffusive regime with increasing Reynolds number when we normalize time by $\tau_{\eta}$.
  However, due to the dependence of 
$\tau_{\eta}$ on Reynolds number, normalizing by the Kolmogorov time confuses a physical interpretation of the diffusion curves.  Using $\mathsf{T_E}$ instead allows for a close comparison of the diffusion curves, which collapse to show a universal trend (see  figure~\ref{figspdiffusion}).  At short times, our single-particle diffusion curves demonstrate the expected ballistic scaling with time to the power of two.  At long times, a diffusive regime is expected; for homogeneous isotropic turbulence the curves scale as a power of one in this regime.

\begin{figure}
\resizebox{2.8in}{!}{\includegraphics{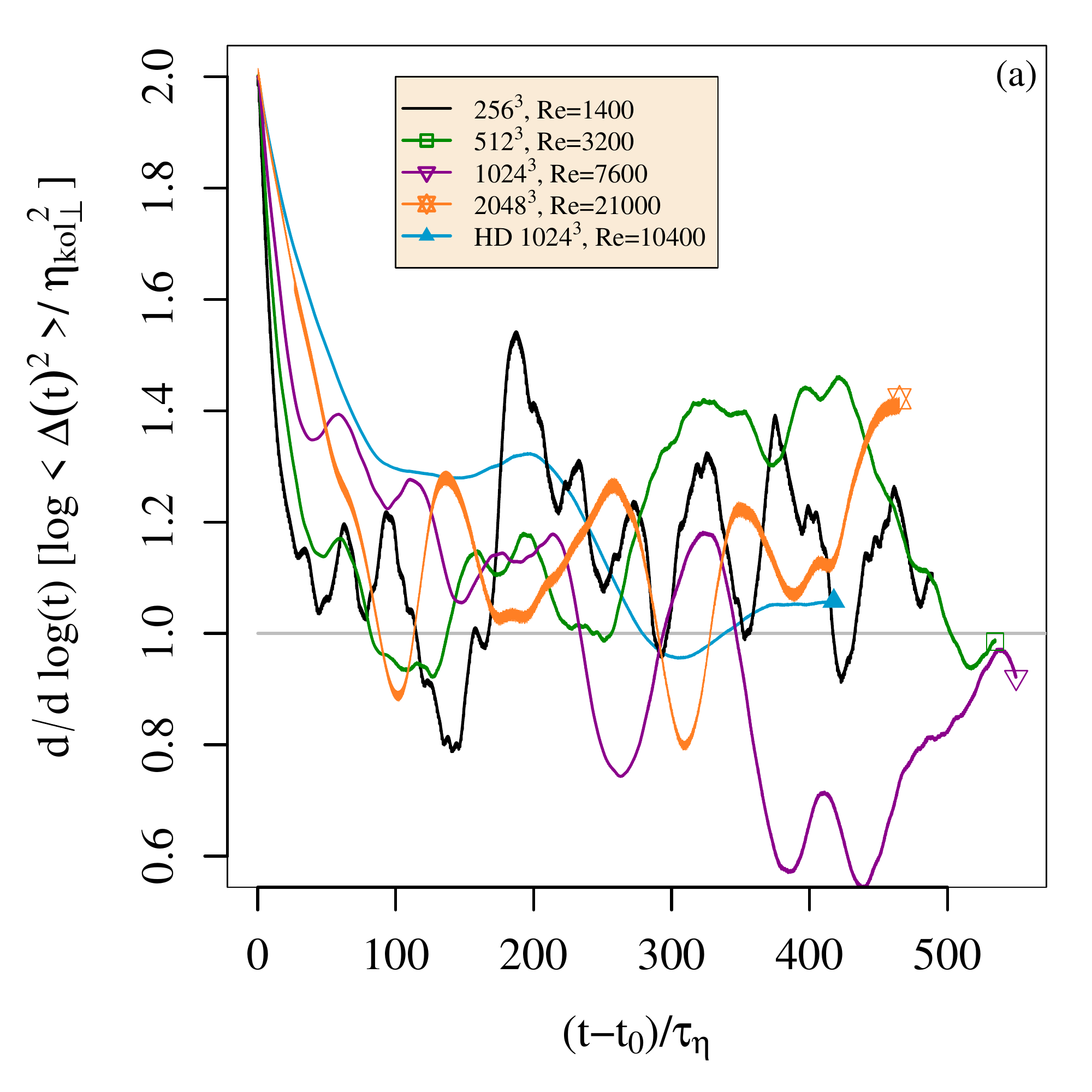}}
\resizebox{2.8in}{!}{\includegraphics{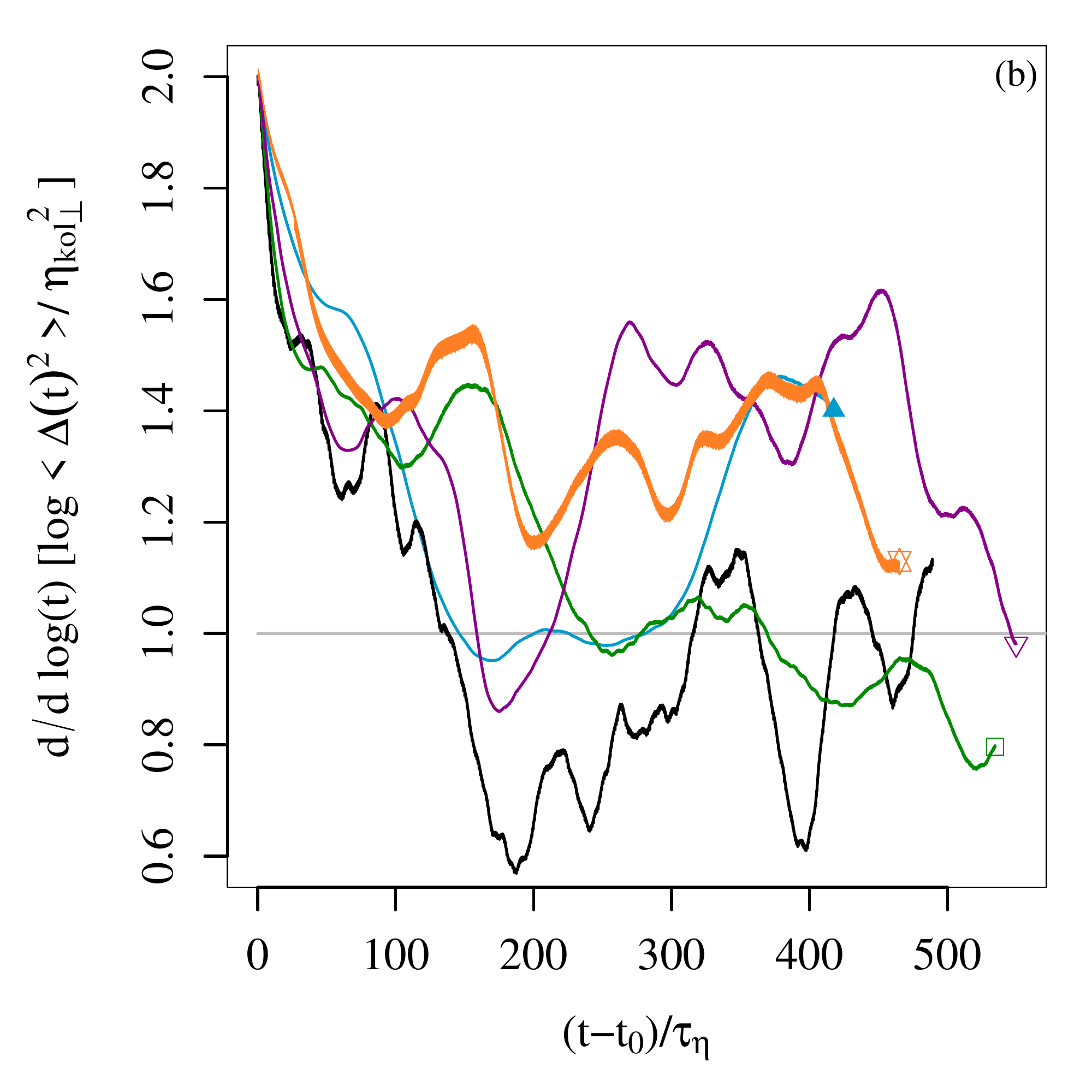}}
\caption{Derivative of the log of $\langle \Delta^2(t) \rangle$ as in figure~\ref{figspdiffusion} for (a) separation perpendicular to the mean magnetic field and (b) separation aligned with the mean magnetic field.  Each diffusion curve represents an average over at least three independent initial times. Distance is measured in units of the Kolmogorov length scale $\eta_{\mathsf{kol},\perp}$ and time is measured in units of the Kolmogorov time scale $\tau_{\eta}$. A grey line indicates the theoretical prediction that the diffusion curve scales linearly with time.
\label{figspdiffusion2}
}
\end{figure}

A close examination of the diffusive scaling in our simulations (see figure~\ref{figspdiffusion2}) reveals oscillations in the log derivative curves in the diffusive regime.  This derivative is calculated using a simple forward Euler method.   Fluctuations are present in homogeneous isotropic turbulence, however the oscillations that we observe in anisotropic MHD turbulence tend to be more regular and noisier.  In the perpendicular direction a scaling with an average value close to unity has been established.  In the parallel direction, this is also true for the lower Reynolds number simulations.  The highest Reynolds number simulation continues to have an average slope that is clearly steeper than one throughout the simulation time; at no point is this slope as low as one.  This measurement of parallel diffusion is calculated parallel to $B_0$ rather than parallel to the local mean field;  particularly in the present case where $B_0$ is approximately equal to the fluctuations in magnetic field, that difference may be significant, so that a deeper interpretation of this anisotropy becomes difficult.

\section{Lagrangian statistics for particle-pairs \label{secresults2}}

Large-scale flow features, in the present case Alfv\'enic fluctuations \citep[e.g.][]{howes2015inherently}, can dominate Lagrangian statistics based on single particles.  We therefore focus our attention on diagnostics based on pairs of Lagrangian tracer particles to expose characteristics of the smaller scales of turbulence. 

\subsection{Lagrangian two-particle dispersion \label{secdispersion}}

The most commonly examined pair statistic is two-particle dispersion.  Two-particle dispersion is the separation of a pair of particles relative to each other, and is usually expressed as a mean-square displacement.  The separation of a pair of Lagrangian tracer particles labeled $i$ and $j$ respectively is simply $\vec{\xi} = \vec{r}_i(t) - \vec{r}_j(t)$ where $\vec{r}_i$ is the vector position of particle $i$ in three-dimensional space.  Dispersion is typically calculated as $\langle (\vec{\xi} - \vec{\xi}_0)^2 \rangle$ where the angular brackets denote an average over all particle pairs that have an initial separation $\vec{\xi}_0$.  At time $t=t_0$, the quantity $\langle (\vec{\xi} - \vec{\xi}_0)^2 \rangle$ is identically zero.  Three subranges are theoretically predicted to exist for isotropic hydrodynamic turbulence:
\begin{eqnarray}
\langle (\vec{\xi} - \vec{\xi}_0)^2 \rangle \sim 
 \begin{cases}
    t^2, & \text{ballistic regime} \\
    t^3, & \text{Richardson regime}  \\
    t,  & \text{diffusive regime. } 
  \end{cases} 
\end{eqnarray}
These three predictions are relevant to short separation times, intermediate separation times, and long separation times, respectively. 
The ballistic regime and diffusive regime have been theoretically motivated, and confirmed by simulations and experiments for hydrodynamic turbulence; simulations have also confirmed that these two regimes exist for isotropic MHD turbulence.  The Richardson scaling
is a plausible prediction for high-Reynolds-number hydrodynamic turbulence \citep[for example see figure 5 of ][]{bourgoin2015turbulent}, and assumes that the initial separation of the pair of particles is arbitrarily small.  For details of the derivations of these scaling laws and further work to improve them we refer to the reviews of  \citet{salazar2009two}, and \citet{sawford2001turbulent}.

\begin{figure}
\resizebox{2.8in}{!}{\includegraphics{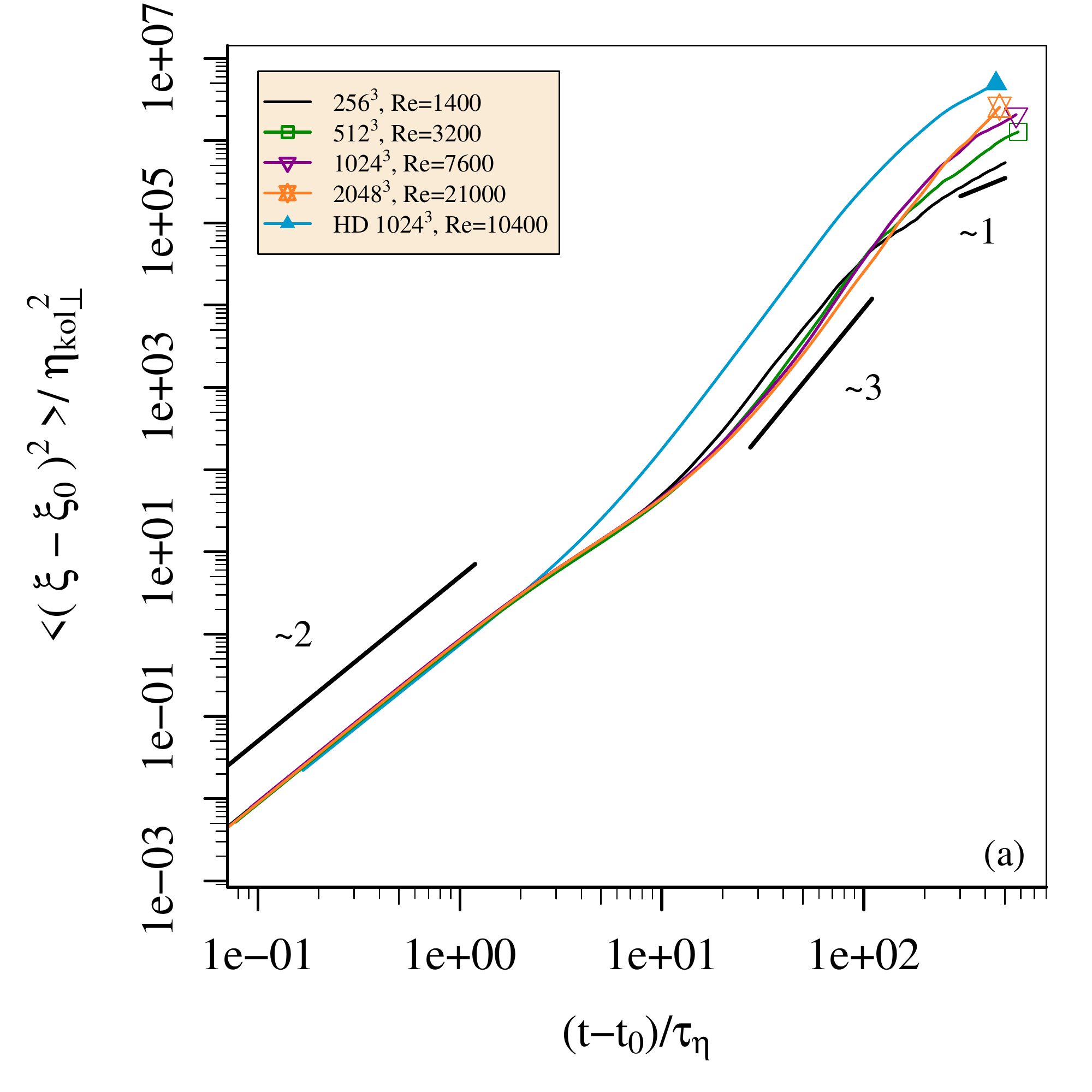}}
\resizebox{2.8in}{!}{\includegraphics{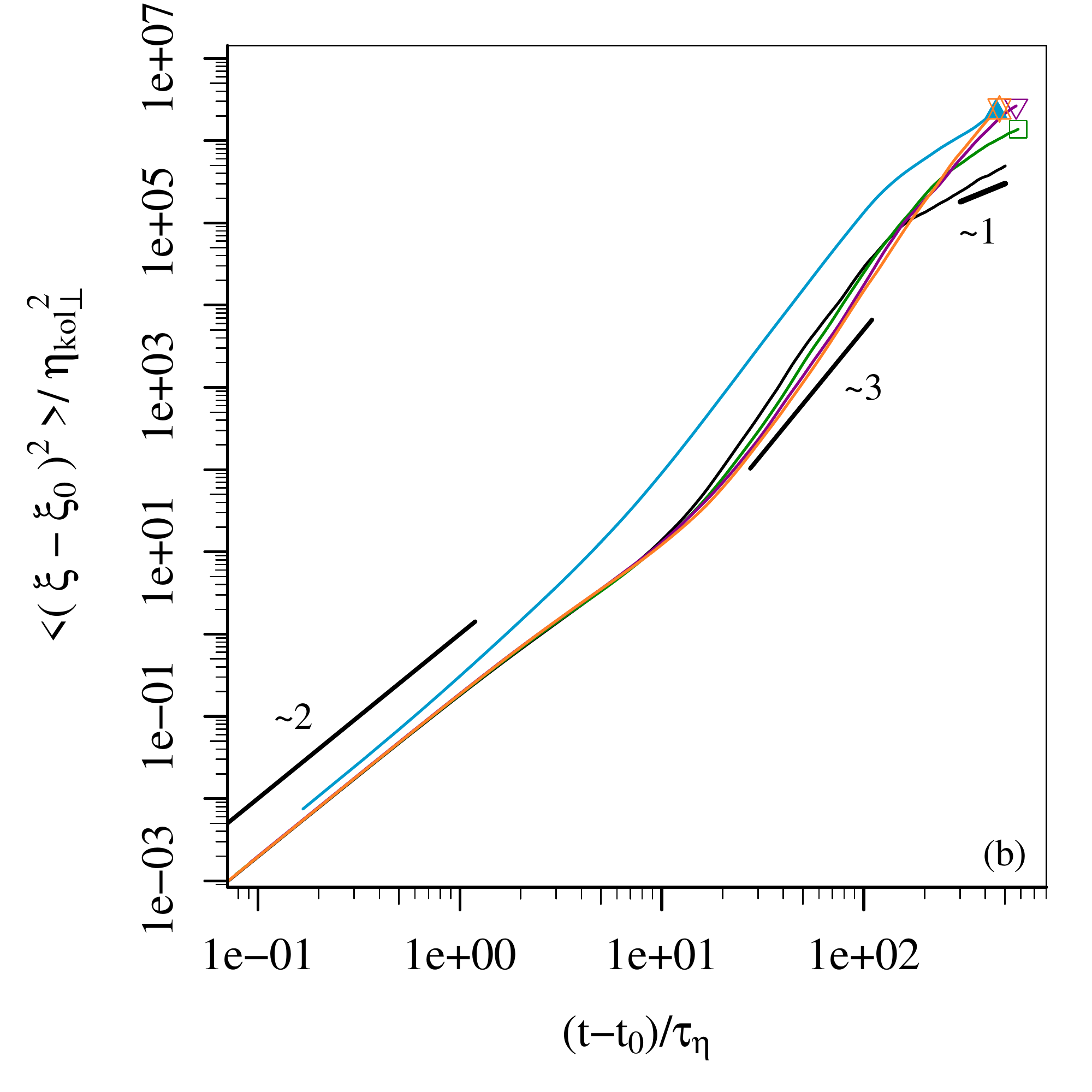}}
\caption{Average square separation of particle pairs initially separated by $\xi_0=2 \eta_{\mathsf{kol},\perp}$ for (a)  perpendicular pairs and separation measure perpendicular to the mean magnetic field, and (b) 
parallel pairs and separation measured in the direction
aligned with the mean magnetic field.  Reference scalings for the ballistic regime, Richardson regime, and diffusive regime are included as straight black lines, with the scaling exponent labeled.
\label{figpairdisp}
}
\end{figure}

We examine dispersion curves for pairs of particles that are initially separated in either the direction perpendicular or parallel to the mean magnetic field;  we call these `perpendicular pairs' and `parallel pairs' respectively.
We also calculate the distances that the particle pairs separate in the perpendicular direction and the parallel directions.  Because a pair of particles has an initial separation that is small compared to the large-scale flow features, two-particle statistics avoid the influence of large-scale sweeping.  The most relevant time-scale for measuring two-particle dispersion is therefore the Kolmogorov time-scale $\tau_{\eta}$ rather than the large-eddy turnover time $\mathsf{T_E}$. 
Dispersion curves for pairs initially separated by $2 \eta_{\mathsf{kol},\perp}$ are displayed in figure~\ref{figpairdisp}.  
This is the closest initial separation for the particle pairs that we have followed; larger initial separations and directions also display a scaling regime with a clear slope but the transitions are less sharp.
The dispersion curves of our four simulations appear nearly identical at short times until approximately $10 \tau_{\eta}$, a range in time corresponding roughly to the ballistic regime; the length of this range appears similar in parallel and perpendicular directions.  During the ballistic regime, faster relative dispersion is observed in the perpendicular direction shown in figure~\ref{figpairdisp}(a)  than in the parallel direction shown in figure~\ref{figpairdisp}(b).
 At long times approaching $400 \tau_{\eta}$, the scaling of the dispersion curves is steeper than one.  For the three simulations with lowest Reynolds number (sim. no. 1--3), the single particle diffusion curves have established a clear diffusive scaling at these late times, characterized by a mildly superdiffusive trend.

Following \citet{salazar2009two}, we refer to the middle range of scales, between approximately $20 \tau_{\eta}$ and $100 \tau_{\eta}$, as the \emph{inertial subrange} without implying that the Richardson scaling or any other scaling with time is a correct prediction for anisotropic MHD turbulence.
For simulations 1 and 2 in table~\ref{simsuma}, which have lower Reynolds numbers, the dispersion curves display a mild curvature that prevents the clear definition of a scaling during the inertial subrange of time scales.  For the higher Reynolds number simulations numbered 3 and 4, the dispersion curves flatten during the inertial subrange and approach a scaling prediction that is clearer.  In the perpendicular direction, for the pairs with initial separation $\xi_0=2 \eta_{\mathsf{kol},\perp}$ shown in figure~\ref{figpairdisp}, this scaling approaches 3, a number reminiscent of the Richardson prediction.  In the parallel direction, the equivalent scaling appears to be steeper than 3.

\begin{figure}
\resizebox{2.8in}{!}{\includegraphics{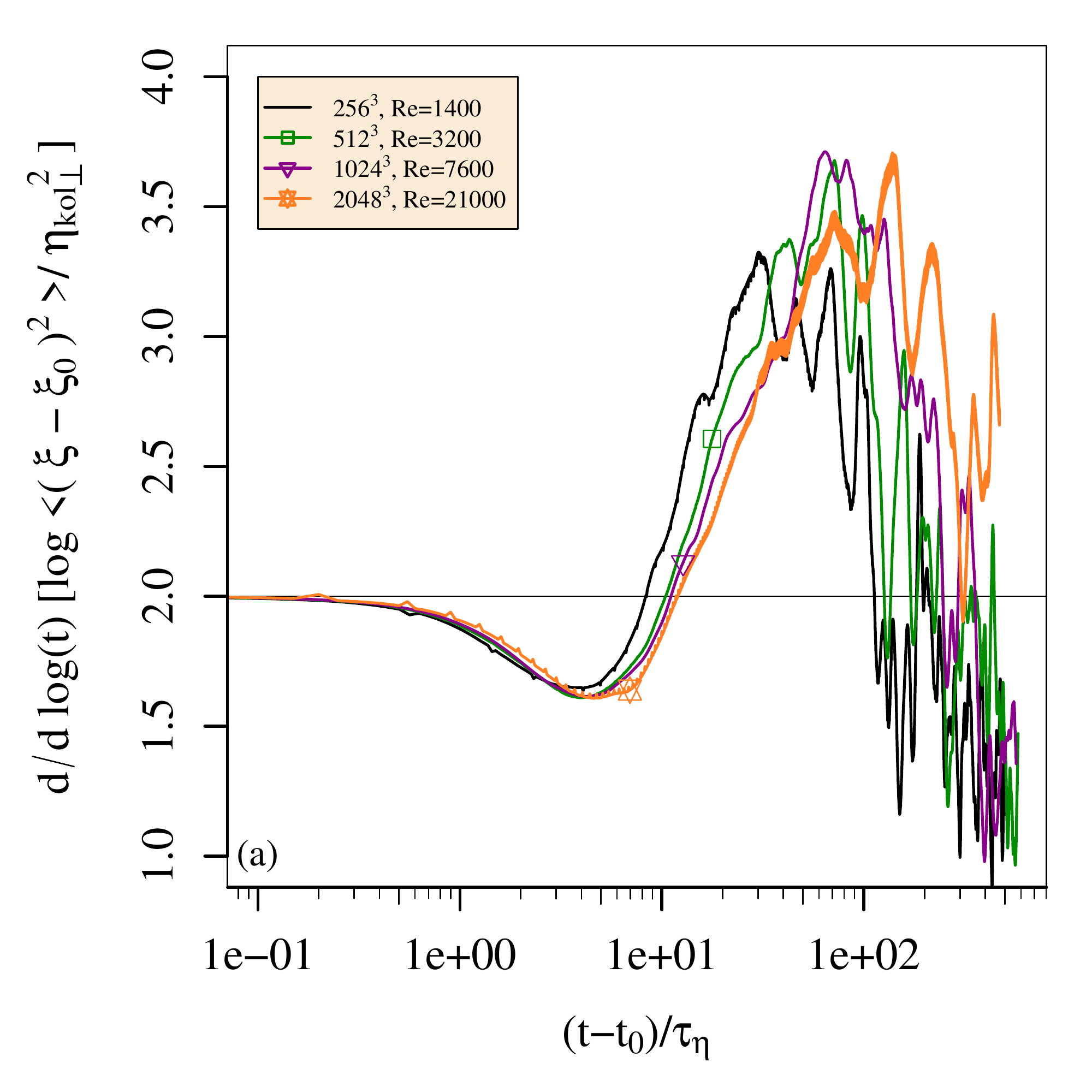}}
\resizebox{2.8in}{!}{\includegraphics{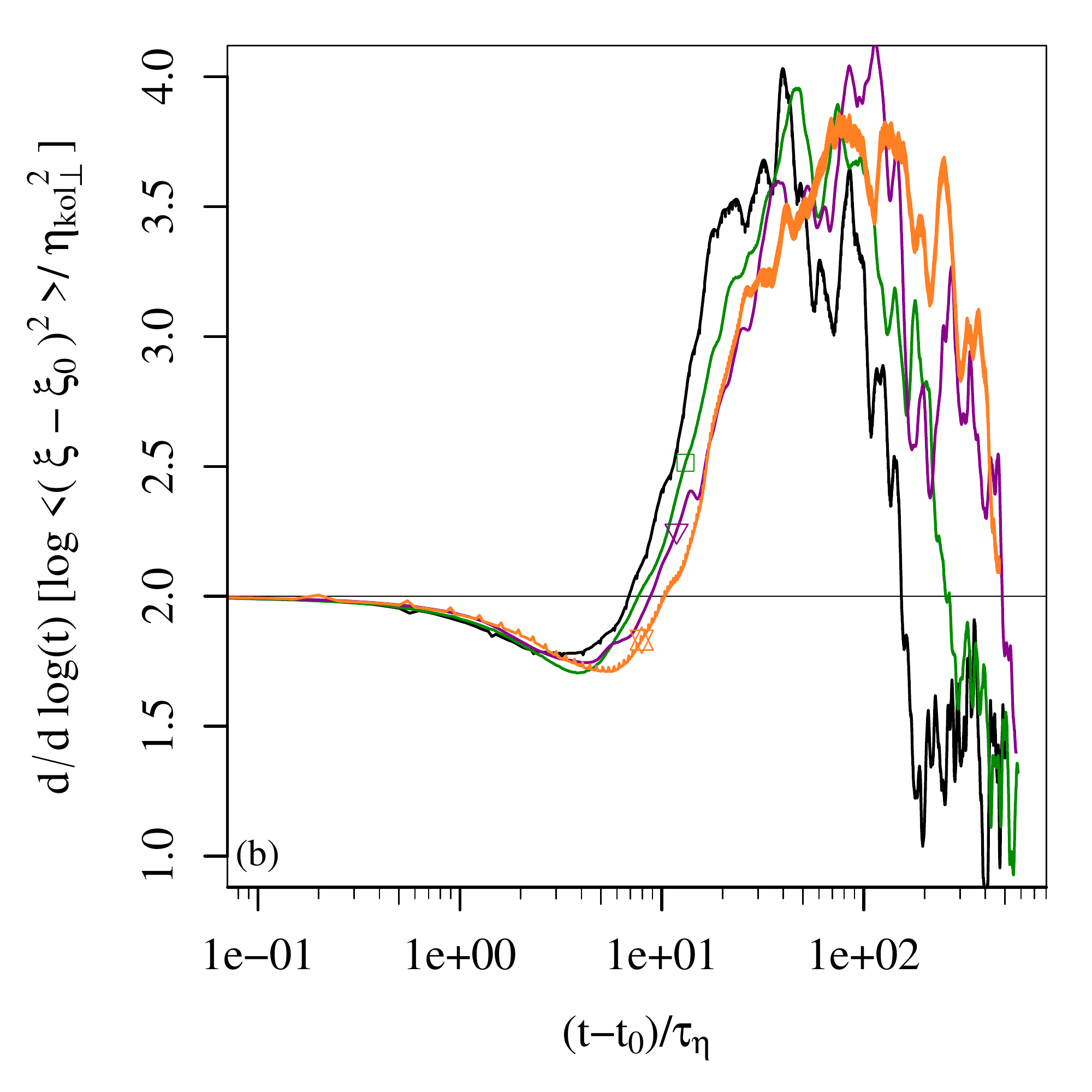}}
\caption{Derivative of the log of the average square separation of particle pairs for (a) perpendicular pairs and separation measured perpendicular to the mean magnetic field,
and (b) parallel pairs and separation measured in the direction
aligned with the mean magnetic field, as in figure~\ref{figpairdisp}.  The initial separation of particle pairs is $\xi_0=2 \eta_{\mathsf{kol},\perp}$.   A grey line indicates the theoretical prediction that the dispersion curve scale with the square of time at early times.
\label{figsnovistwodlog}
}
\end{figure}

We examine the log derivative, to quantitatively analyze these scalings (see figure~\ref{figsnovistwodlog}).  Here the derivative of the log is calculated using a simple forward Euler method, because such a two-point method allows the very early behavior to be seen most clearly.  The log derivative shows an initial slope of 2 during the ballistic regime for both perpendicular and parallel results.
The inertial subrange and diffusive regime are both characterized by chatter in the log derivative.  During the inertial subrange, between roughly $20 \tau_{\eta}$ and $100 \tau_{\eta}$, the log derivative of the perpendicular dispersion curve for perpendicularly separated pairs ranges between 2.3 and 3.7, while the parallel dispersion of pairs separated in the parallel direction ranges between approximately 3.0 and 4.0.  The range over which these derivatives chatter provides an estimate of the uncertainty for the scaling of the dispersion curves.  Some of the noisiness here results from the fact that our two-particle statistics are constructed from a single initial time.  During the diffusive regime both perpendicular and parallel dispersion curves decay to values between approximately 1.0 and 2.0.  Simulations 1 and 2 have an average scaling greater than 1 and less than 1.5 between $300 \tau_{\eta}$ and $400 \tau_{\eta}$, corresponding to a slightly superdiffusive separation.

\begin{figure}
\resizebox{2.8in}{!}{\includegraphics{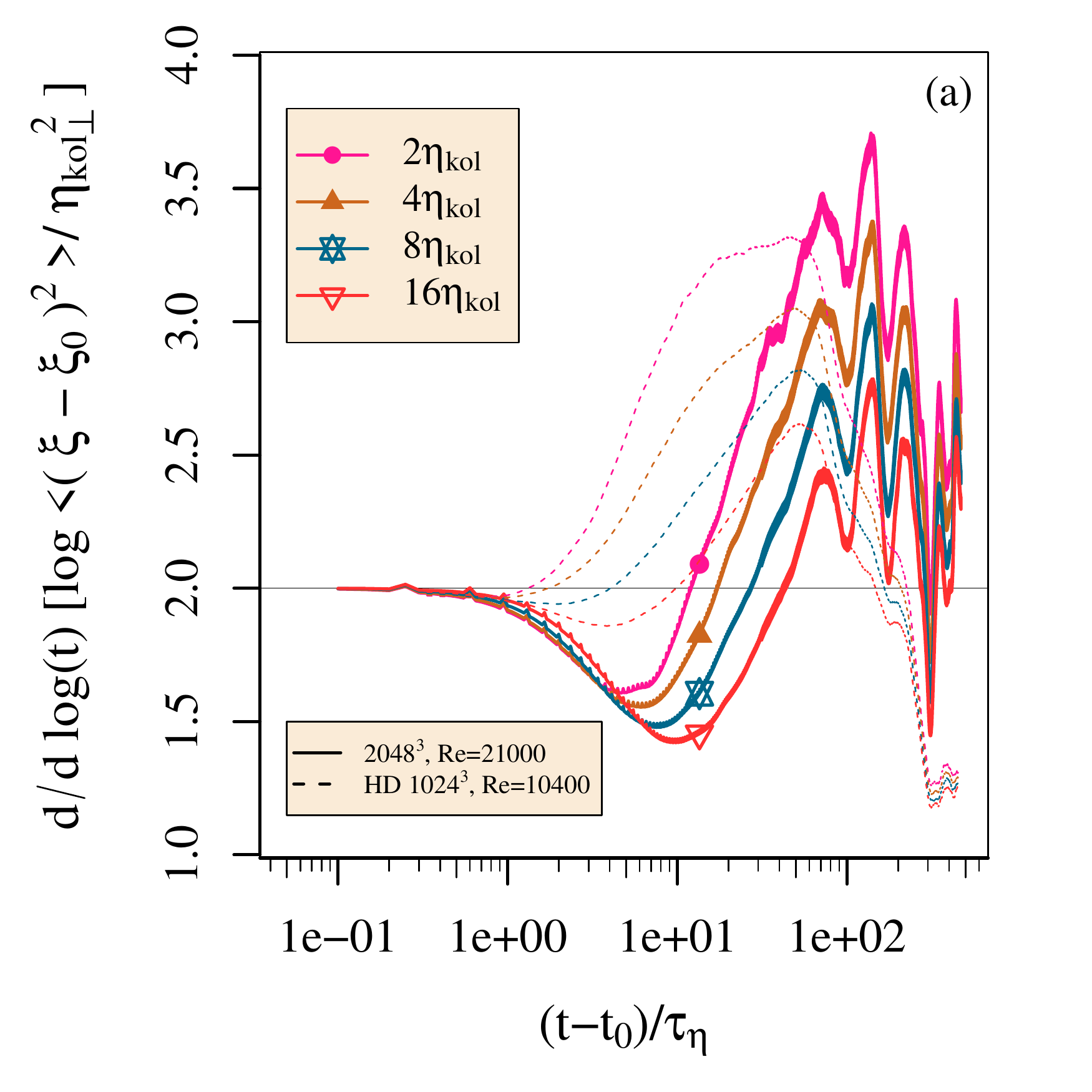}}
\resizebox{2.8in}{!}{\includegraphics{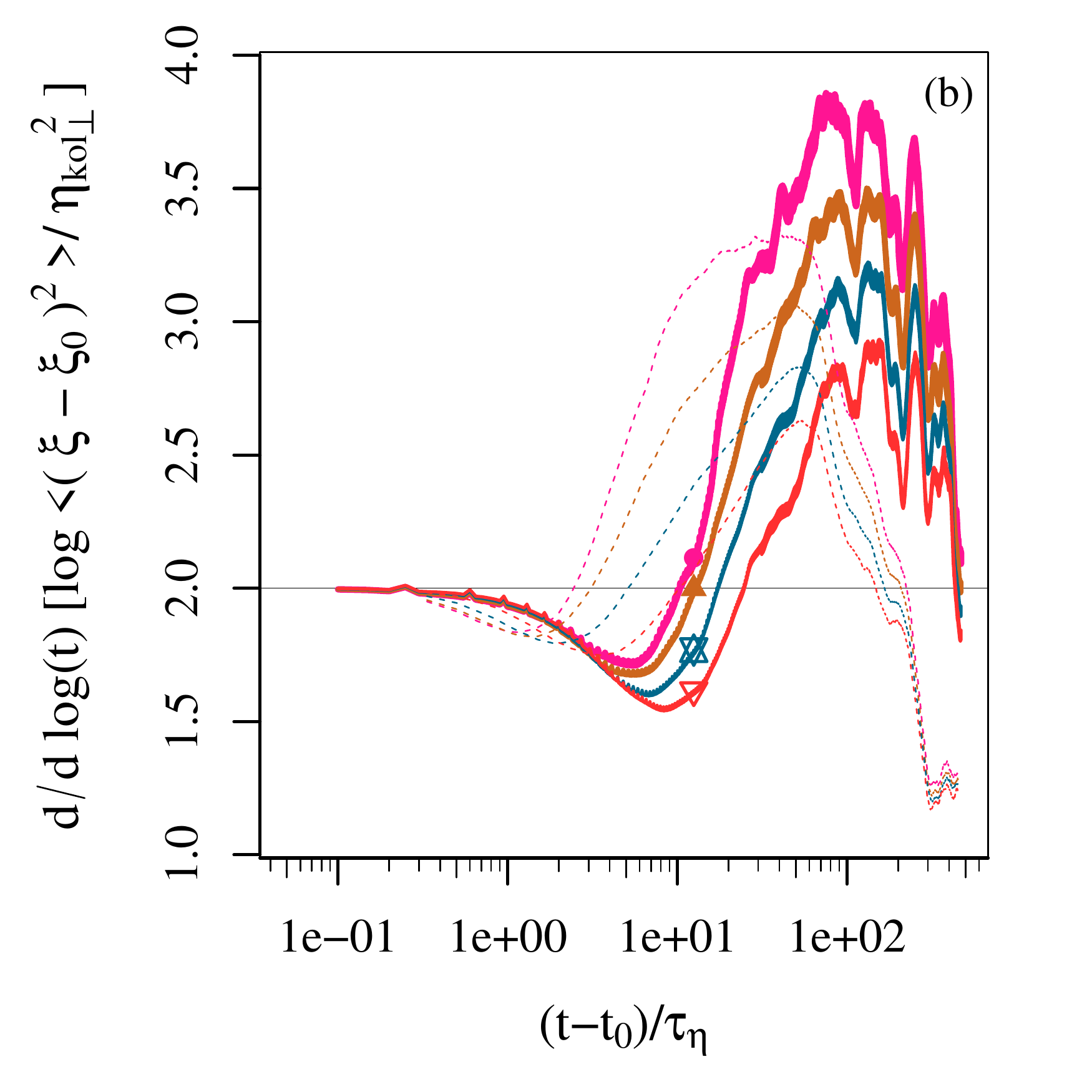}}
\caption{Log derivative of the average square separation of particle pairs in the direction perpendicular to the mean magnetic field for (a) perpendicular pairs and separation measured perpendicular to the mean
magnetic field, and (b) parallel pairs and separation measured in the
direction aligned with the mean magnetic field.
  Each line is labeled by the initial separation distance of the particle pairs.
Data from simulation 4, described in table~\ref{simsuma}.  Equivalent curves from simulation 3H are shown as dashed lines in the background for comparison.   A grey line indicates the theoretical prediction that the dispersion curve scale with the square of time at early times.
\label{figsnovispairs}
}
\end{figure}

As we have discussed, for particle pairs that are initially separated by $\xi_0=2 \eta_{\mathsf{kol},\perp}$, the separation in the perpendicular direction appears close to the Richardson prediction of a scaling of 3.  Calculating the log derivative for particle pairs with different initial separations $\xi_0$ in simulation 4 (see figure~\ref{figsnovispairs}) clarifies that this scaling is indeed dependent on $\xi_0$ \citep[as discussed by][]{biferale2005lagrangian}. 
This figure also provides a comparison to the hydrodynamic simulation 3H.  For anisotropic MHD turbulence, the log derivative reveals larger changes between the ballistic and the inertial subranges than in the hydrodynamic case.  For each group of particle pairs with the same initial separation, there is first a dip before or near $\tau_{\eta}$  indicating a temporary slowing down of dispersion. This dip is followed by a series of peaks as the pair separation enters the inertial subrange; since the hydrodynamic simulation has a single smooth peak, the multiple peaks and the chatter during this period are likely to result from Alfv\'enic fluctuations.  The maximum value of the log derivative for anisotropic MHD turbulence is a higher value than the hydrodynamic case.   Thus, although the dispersion curves for simulation 4 point toward a Richardson-like scaling regime for the perpendicular dispersion, they do not clearly confirm Richardson scaling for anisotropic MHD turbulence.  The log derivative curves indicate a larger slope is achieved during the inertial subrange for a smaller initial separation $\xi_0$.  Therefore in the limit where $\xi_0$ becomes arbitrarily small,  we expect that the slope of the dispersion curves will be greater than 3 in both the perpendicular and parallel directions.  We thus also expect that corrections to a Richarson-like prediction are needed for the case of anisotropic MHD turbulence.

\begin{figure}
\resizebox{2.8in}{!}{\includegraphics{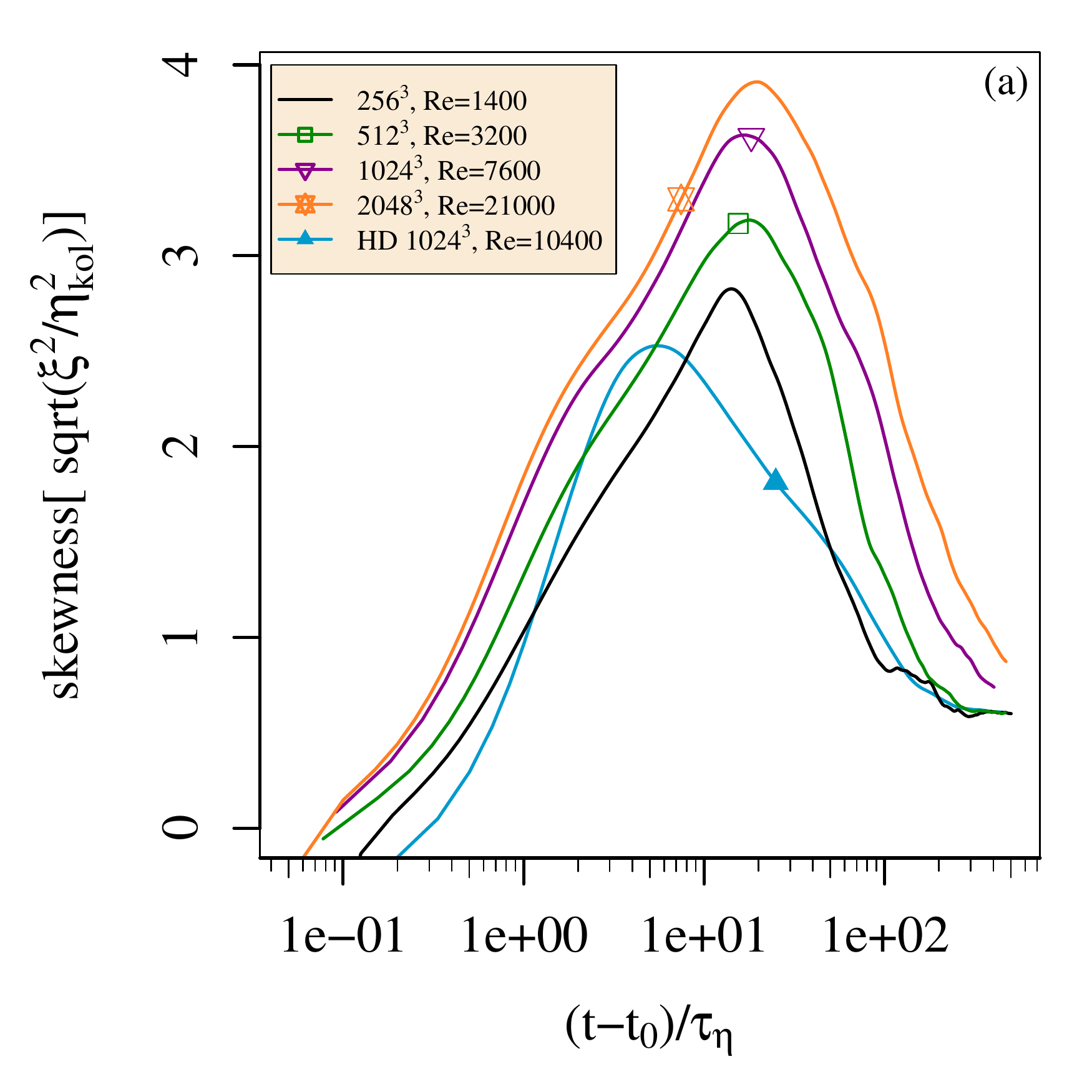}}
\resizebox{2.8in}{!}{\includegraphics{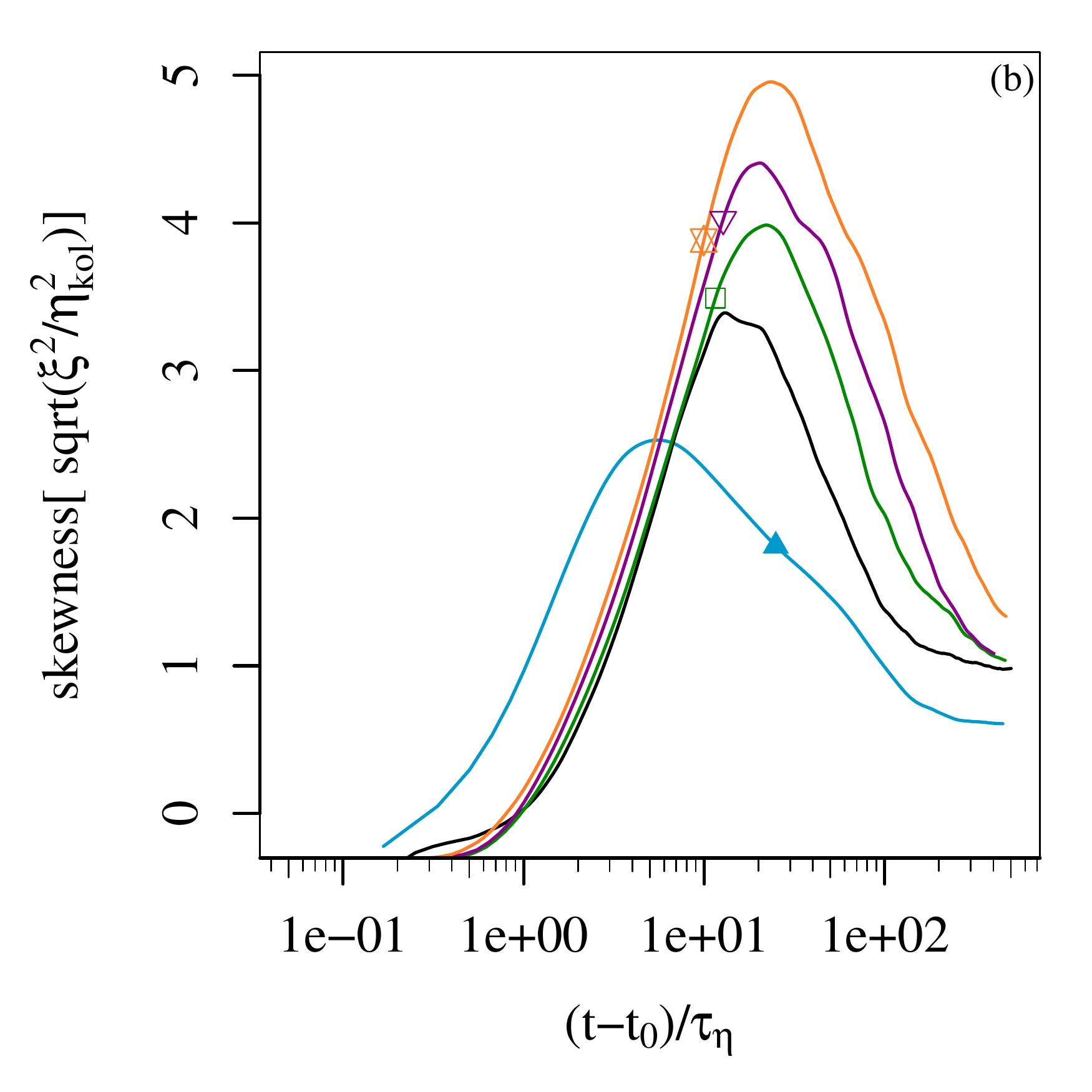}}
\caption{Skewness of the particle-pair separations for (a) perpendicular pairs with separation measured in the perpendicular direction, and 
(b) parallel pairs with separation measured in the aligned direction.  For comparison with isotropic hydrodynamic turbulence, we provide the blue line from simulation 3H.  The initial separation of the particle pairs is $4 \eta_{\mathsf{kol},\perp}$ in each simulation.
\label{figsnovispairsskewness}
}
\end{figure}

It has been shown \citep[e.g.,][]{yeung2004relative}, that the intermittency of particle-pair dispersion is larger in higher Reynolds number simulations of isotropic hydrodynamic turbulence.  In hydrodynamic turbulence, increasing the Reynolds number leads to a stronger
intermittency at the small scales; the separation of particle pairs provides a convenient measure for this, since they sample the velocity field on a length scale comparable to their separation distance.
The skewness, a normalized third moment, indicates the asymmetry of the wings of a distribution; it is therefore one indicator of intermittency, which can also be observed in other high-order moments.  A larger skewness of the distribution of particle-pair separations indicates the importance of the extremes of dispersion, and the relative importance of pairs of particles that separate much faster than the average.  In our simulations of anisotropic MHD turbulence, the skewness of the separation distance provides a particularly clear result that demonstrates the Reynolds number dependence (see figure~\ref{figsnovispairsskewness}).   In the figure, the particle pairs are initially separated by $4 \eta_{\mathsf{kol},\perp}$, so that this diagnostic can be compared with figure 13 of \citet{yeung2004relative}. We also include the results from our simulation 3H in this figure to provide a direct comparison with the isotropic hydrodynamic case.   We find that anisotropic MHD turbulence develops a larger skewness of the particle-pair separations than isotropic hydrodynamic turbulence, even for similar Reynolds number simulations. 

At early times, the values of the skewness of particle-pair separations are small and negative, but quickly become positive.  The skewness rises more rapidly at earlier times for perpendicular pairs than for parallel pairs.   This difference may be due to the presence of Alfv\'enic fluctuations, oscillations on large-scale temporal and spatial scales, which become elongated in the direction of the mean magnetic field.  The initialization of particle pairs allows some pairs to reside entirely inside such flow structures.  Other particle pairs saddle the boundary of a flow structure, with one particle inside and another particle outside; those pairs tend to separate more rapidly than the average.  Because of the elongation of flow structures, pairs that saddle such boundaries are more likely to have an initial separation in the perpendicular direction.  

As the pairs of particles move further apart in space, they experience a decorrelation between their local velocities.  Eventually the rate of separation of these early fast-separating pairs slows, and the skewness reaches a peak. The peaks in skewness for our anisotropic MHD simulations occur at a time of approximately $10 \tau_{\eta}$, a later time than for our isotropic hydrodynamic turbulence simulation.  This peak is positioned near the point of transition between the ballistic regime and the inertial subrange.  We observe no clear difference in the placement of this peak with Reynolds number, or between the anisotropic directions.
 The peak in skewness is higher for larger Reynolds number,  and is also higher for parallel pairs than for perpendicular pairs.   
 The difference in the height of the peak indicates that intermittency is more intense along the direction of the mean magnetic field, as well as more intense for higher Reynolds number.  This difference between the parallel and perpendicular directions suggests that the current sheets that define anisotropic MHD turbulence could be responsible for more frequent large particle separations in this setting.

\subsection{Two-particle velocity statistics \label{secresults3}}

\begin{figure}
\resizebox{2.8in}{!}{\includegraphics{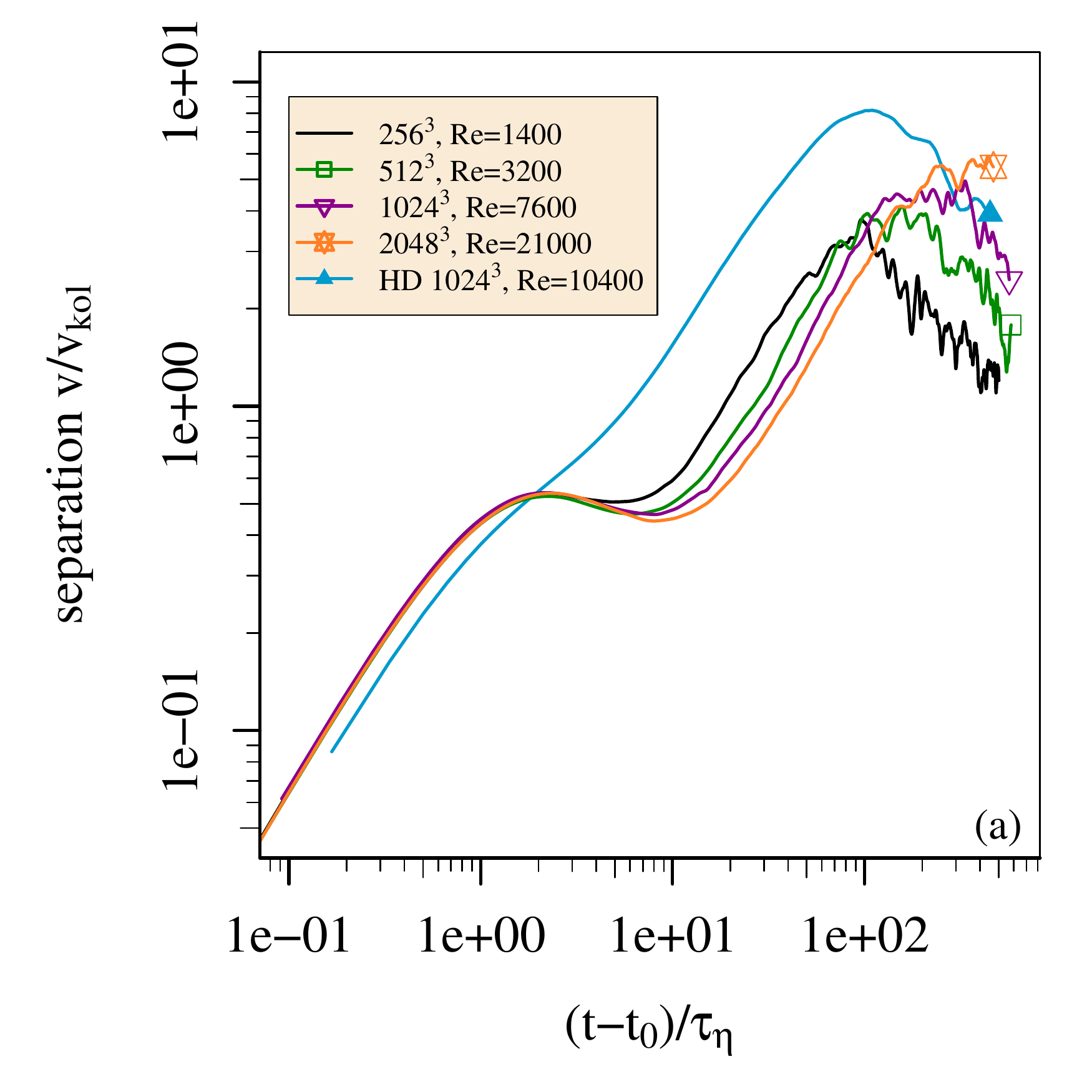}}
\resizebox{2.8in}{!}{\includegraphics{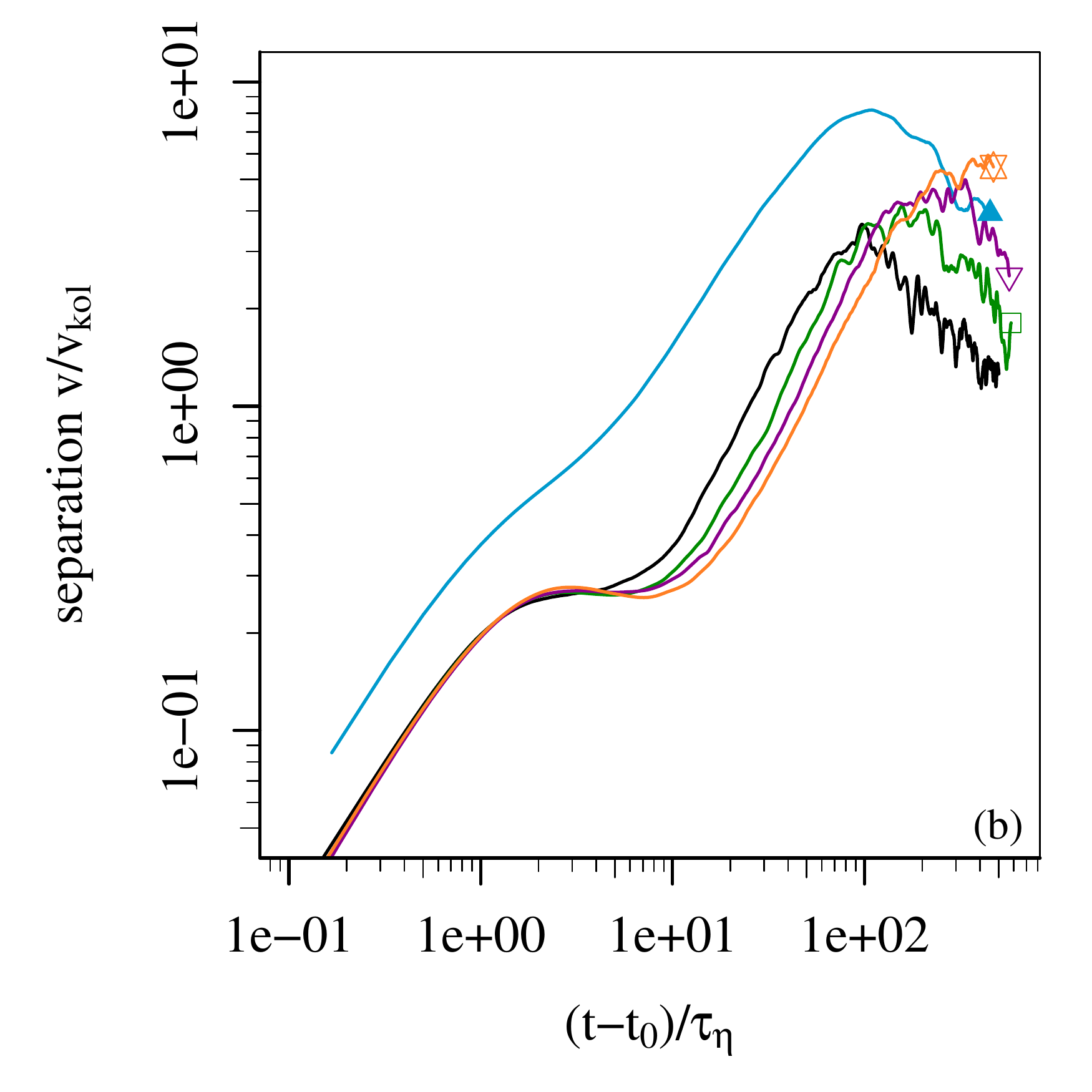}}
\caption{Average separation speed for (a) perpendicular pairs,
and (b) parallel pairs. Time is given in units of the Kolmogorov time-scale $\mathsf{\tau_{\eta}}$.  These pairs of particles are initially separated by $2 \eta_{\mathsf{kol},\perp}$.
\label{figsepvel}
}
\end{figure}

To expose further differences between homogeneous isotropic hydrodynamic turbulence and MHD turbulence, we examine the separation speed, i.e. the projection of the velocity difference experienced by a pair of Lagrangian tracer particles onto the line connecting the particles.
The separation speed has been used to construct stochastic models and other Lagrangian statistics \citep[e.g.][]{sokolov1999two,boffetta2002statistics}, and to examine intermittency and alignment in isotropic hydrodynamic flows \citep{biferale2005lagrangian,yeung2004relative}.   From early times until the end of the inertial subrange, the separation speed is significantly higher for perpendicular pairs than for parallel pairs.  In our simulations, the average separation speed displays a characteristic dip, first identified by \citet{muller2007diffusion}, near the beginning of the inertial subrange of time scales for MHD turbulence (see figure~\ref{figsepvel}).  The isotropic hydrodynamic simulation 3H is provided on the plot for comparison; it does not experience a similar dip. 
This dip occurs between approximately $2 \tau_{\eta}$ and $10 \tau_{\eta}$, a period where pairs of particles have separated sufficiently to sense temporal fluctuations of the velocity field.
At the same time, the slope of the average dispersion curve is changing between the ballistic regime and the inertial subrange; the skewness of the pair separation distance is increasing to a peak.   We find that this characteristic dip in the separation speed is deeper and lasts longer for higher Reynolds number simulations. 

Following the dip, the separation speed rises again as the particle pairs probe ever larger and more energetic fluctuations.
This comes to an end as the pairs reach separation distances comparable to the size of the largest fluctuations in the system.
We find that the separation speed ultimately reaches a higher value for higher Reynolds number simulations. 

As an increasing number of particle pairs reaches the diffusion regime, the average separation velocity begins a period of fluctuating decay. 
At this point the large-scale velocities experienced by the separating particles tend to become decorrelated, resulting in slower, asymptotically 
diffusive separation dynamics. The separation speed is expected to 
level off at a quasi-stationary diffusive separation velocity once most pairs have left the superdiffusive region of pair dispersion.  The fluctuations during this decay period are more noisy for anisotropic MHD turbulence, where Alfv\'enic fluctuations are present at all scales, than for isotropic hydrodynamic turbulence.

\begin{figure}
\resizebox{2.8in}{!}{\includegraphics{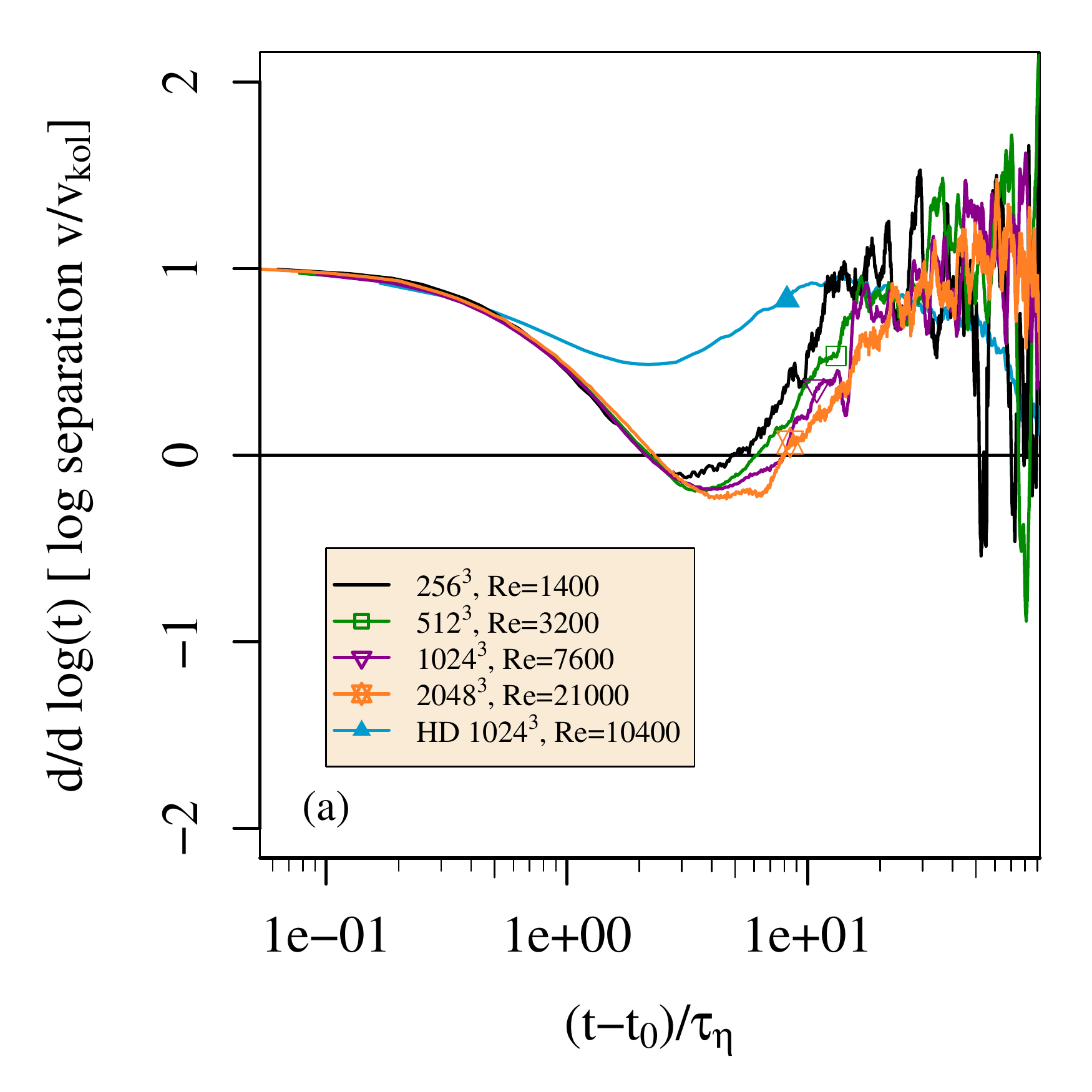}}
\resizebox{2.8in}{!}{\includegraphics{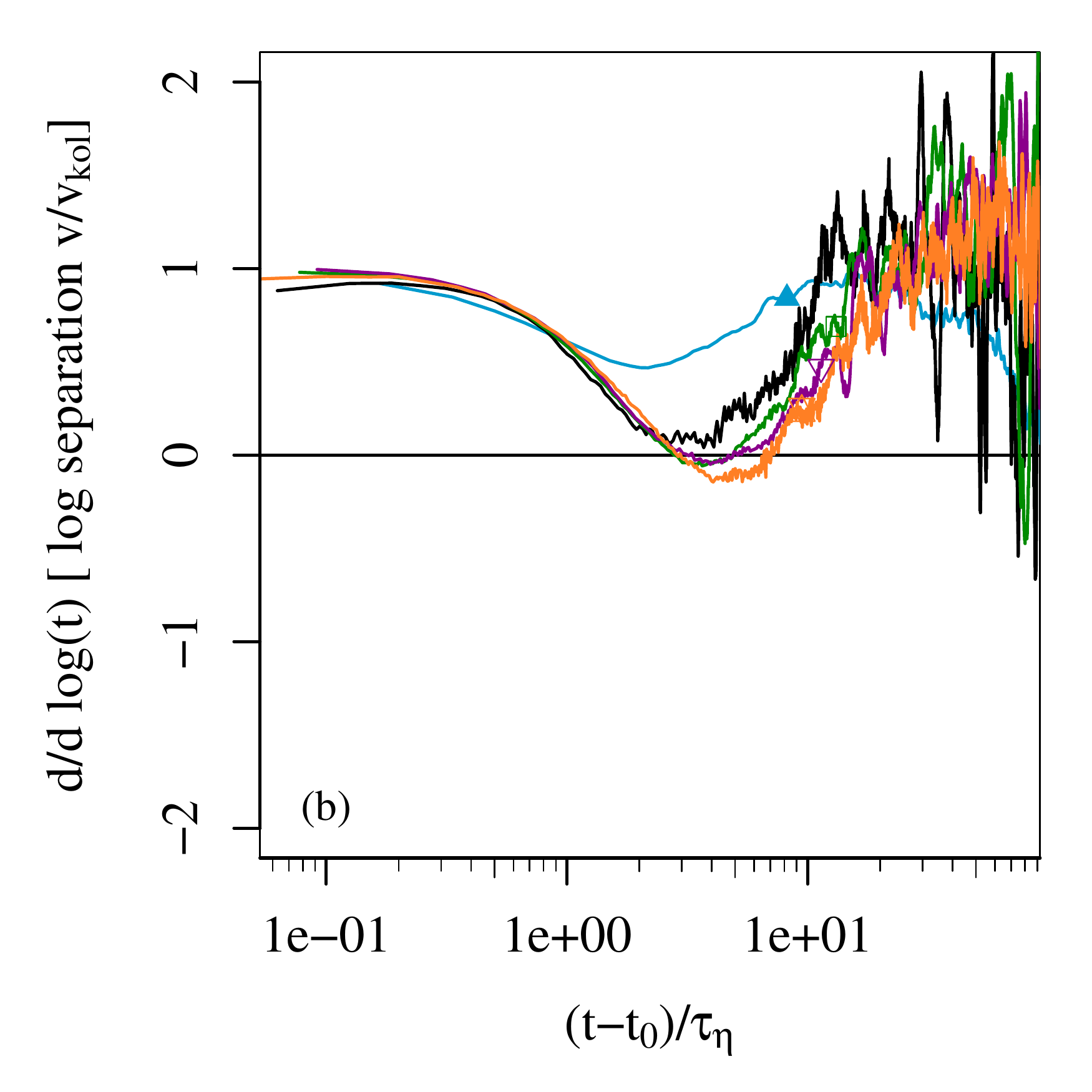}}
\caption{Derivative of the log of the average separation speed for  (a) perpendicular pairs,
and (b) parallel pairs.  These curves correspond to figure~\ref{figsepvel}.
\label{figsepveldlog}
}
\end{figure}

We calculate the log derivative to determine whether a scaling exists with Reynolds number for the separation speed (see figure~\ref{figsepveldlog}).  The log derivative of the separation speed shows that the dip deepens, and is not simply shifted for higher Reynolds number.  Instead, higher Reynolds number causes a milder upward slope at the end of the ballistic regime, and this persists throughout the inertial subrange.  For example at $10 \tau_{\eta}$, the slope is clearly different for different Reynolds number simulations.  This is a consequence of the longer duration of the alignment process for the separation velocity that occurs at higher Reynolds number, as well as the stronger slowing down that resulting from it. 

\subsection{Alignment statistics \label{secresults3}}

\citet{yeung2004relative} explain the behavior of the average separation speed in isotropic hydrodynamic turbulence through an examination of the alignment statistics using
the angle between the relative velocity and separation vector of pairs of particles.   Concepts of alignment also prove useful in explaining differences between isotropic hydrodynamic and MHD turbulence, an idea first explored by \citet{muller2007diffusion}.  Here we expand on those ideas of alignment statistics; we extend the arguments of \citet{muller2007diffusion} to examine the relationship between alignment and Reynolds number in the distinct case of $B_0=1$ anisotropic MHD turbulence.

To quantify how the separation of pairs of particles differs in the anisotropic magnetohydrodynamic case, we examine the average of the cosine of the angle between the relative velocity of pairs of Lagrangian tracer
particles and their separation vector.  Following \citet{yeung2004relative,yeung1994direct}, we describe this angle as the ``alignment angle,'' and designate it with $\beta$.  
  Examining the cosine of the alignment angle is particularly useful because it should be larger when the relative velocity and separation are better aligned.  Recently \citet{malik2021new} have defined a pair diffusion coefficient $K(t)$ to be the average value of the scalar product of the separation vector and the relative velocity, a quantity strongly related to the alignment angle. Their work demonstrates that the alignment angle is integral to scaling laws that can be constructed for the inertial subrange.

\begin{figure}
\resizebox{2.8in}{!}{\includegraphics{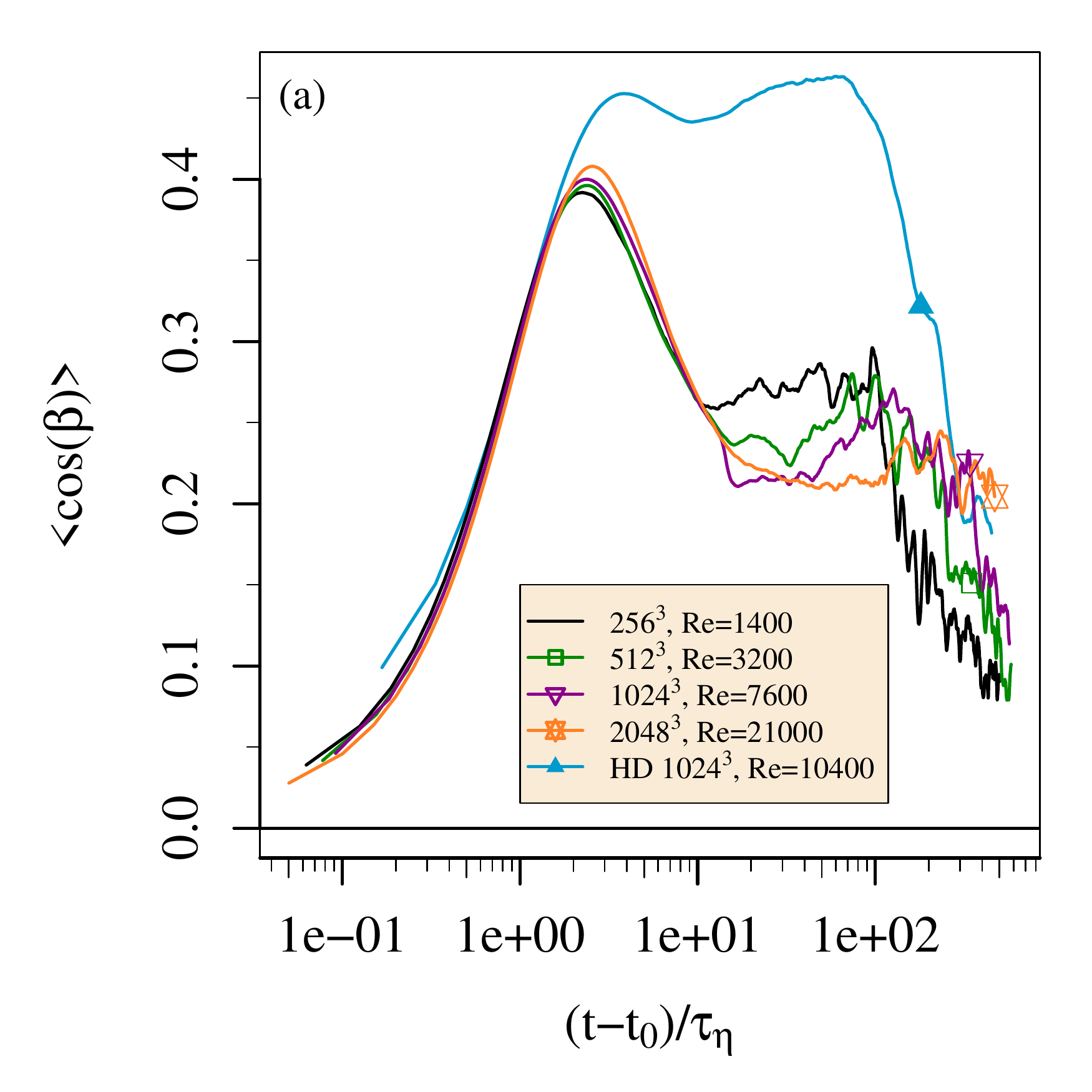}}
\resizebox{2.8in}{!}{\includegraphics{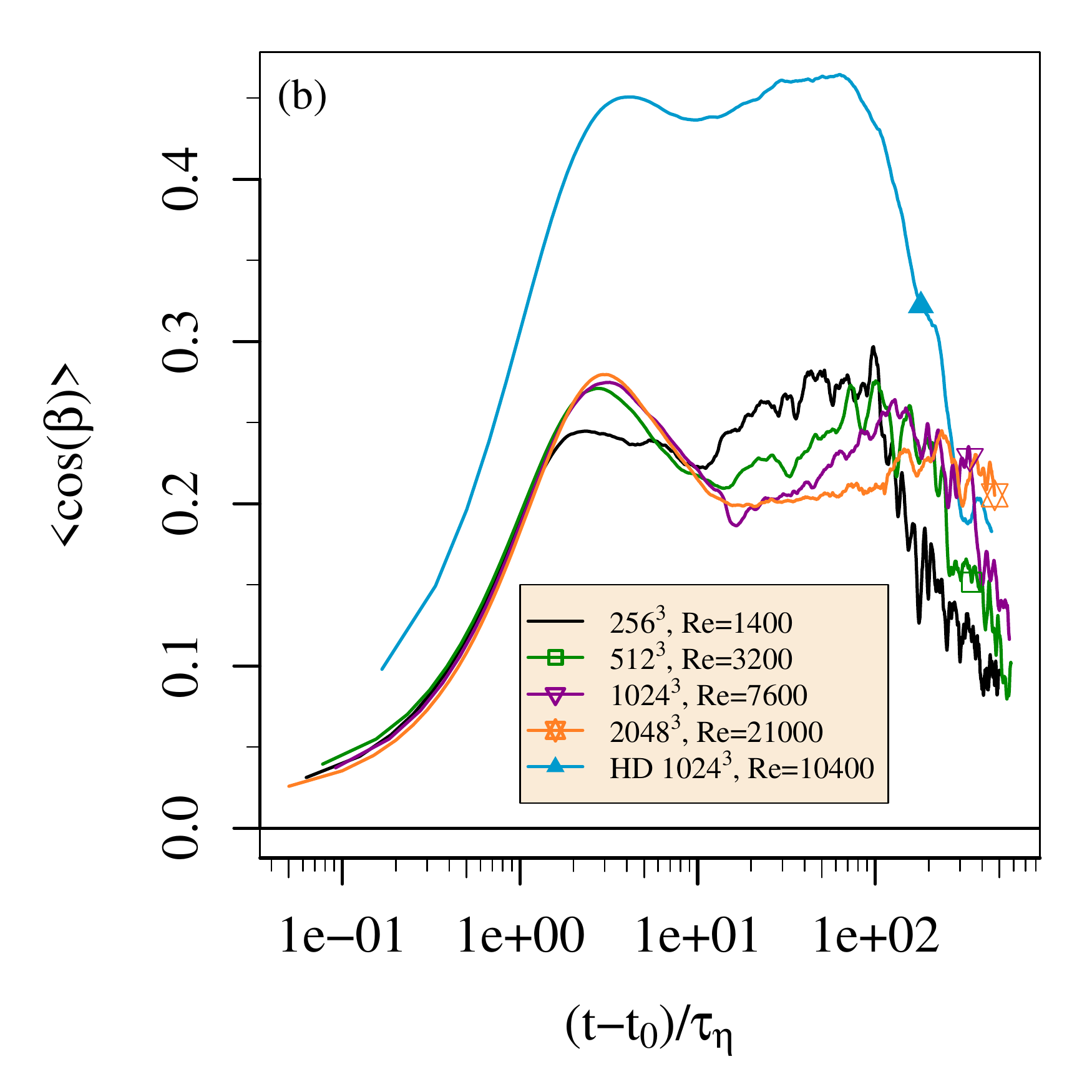}}
\caption{Average cosine of the alignment angle $\beta$.  This average is shown for (a) perpendicular pairs,
and (b) parallel pairs.  These pairs of particles are initially
separated by 2 $\eta_{\mathsf{kol},\perp}$.  Time is given in units of the Kolmogorov time-scale $\mathsf{\tau_{\eta}}$.  
\label{figcosvr}
}
\end{figure}

The average cosine of the alignment angle is clearly different for particles initially separated in the directions parallel and perpendicular to the mean magnetic field.  These diagnostics also display a trend with Reynolds number.  In figure~\ref{figcosvr} the average cosine of the alignment angle grows from approximately zero to a peak at approximately $2 \tau_{\eta}$.  This measure reveals that during these early times, the separation vector of a particle pair tends toward a state in which it is better aligned with the separation velocity.  The growth phase takes place during the ballistic regime of dispersion, as the particles follow the initial velocity of the fluid.
   For perpendicular pairs, the amplitude of the peak
is about 0.4, which is slightly lower, but comparable to, the hydrodynamic case; for parallel pairs, the amplitude of the peak is approximately 0.28.  
 The effect of the particle separations aligning with the relative velocity is larger in the perpendicular direction.     This can be related to the higher separation speed that we observed for perpendicular pairs.

As particle pairs separate further in the flow, they begin to probe the lower end of the inertial subrange of time scales.
The signature of this in the average cosine of the alignment angle is a drop-off between $2 \tau_{\eta}$ and $10  \tau_{\eta}$, after which the average cosine enters a plateau and exhibits noisy behavior.  
The time for the first drop-off, which signifies the loss of initial strong alignment of the particle pairs, correlates with the slow-down in separation velocity noted in figure~\ref{figsepvel}.
   For higher Reynolds number the drop-off is larger, suggesting a link between the loss of alignment and the magnitude of the slow-down in separation speed.   
The isotropic hydrodynamic case is provided in figure~\ref{figsepvel} for comparison; in this case the alignment experiences a small and brief dip, then regains and maintains its initial high alignment throughout the inertial subrange of time scales.  For particle pairs in anisotropic MHD, regardless of their initial separation direction, the value of the average cosine of the alignment angle during the plateau is between 0.2 and 0.3; for isotropic hydrodynamic turbulence, the magnitude is nearly twice as large.  
Finally, at long times corresponding to the diffusive range, the average cosine of the alignment angle again decreases.
For anisotropic MHD turbulence it ultimately drops to about 0.1; for isotropic hydrodynamic turbulence, this value is higher, about 0.2.  In both the inertial subrange and the diffusive regime, the relative dispersion is slower for MHD turbulence, when compared with hydrodynamic turbulence, because of the weaker alignment between the relative velocity and the separation vector.

For MHD turbulence, \citet{muller2007diffusion} demonstrated that the angle between the separation vector and the local mean magnetic field needs to be considered.  We call this the ``magnetic alignment angle'' and designate it with $\gamma$.  The local mean magnetic field is simply the average of the total magnetic field experienced by the pair of Lagrangian tracer particles at their positions at each point in time, which includes contributions of both the fluctuating field and mean field.   
  The average values of $\cos{\gamma}$ should approach zero for a perfectly isotropic turbulent flow.
  In an anisotropic flow, where the initial magnetic alignment angle needs to be taken into account, this is not the case.
For  perpendicular pairs, 
alignments perpendicular to the local magnetic field have an increased probability because the local magnetic field tends to be dominated by the mean magnetic field.   Thus the initial value for $\langle \cos{\gamma} \rangle$ is close to the zero.  For parallel pairs, the initial value for $\langle \cos{\gamma} \rangle$ is finite because the separation vector between particles consistently points in the direction antiparallel to the mean magnetic field.  Regardless of the initial separation direction, as the pairs of particles evolve in time, $\langle \cos{\gamma} \rangle$ approaches zero because the dependency on the initial alignment state disappears.     The average thus does not display a clear Reynolds number dependence in our simulations.  

\begin{figure}
\resizebox{2.8in}{!}{\includegraphics{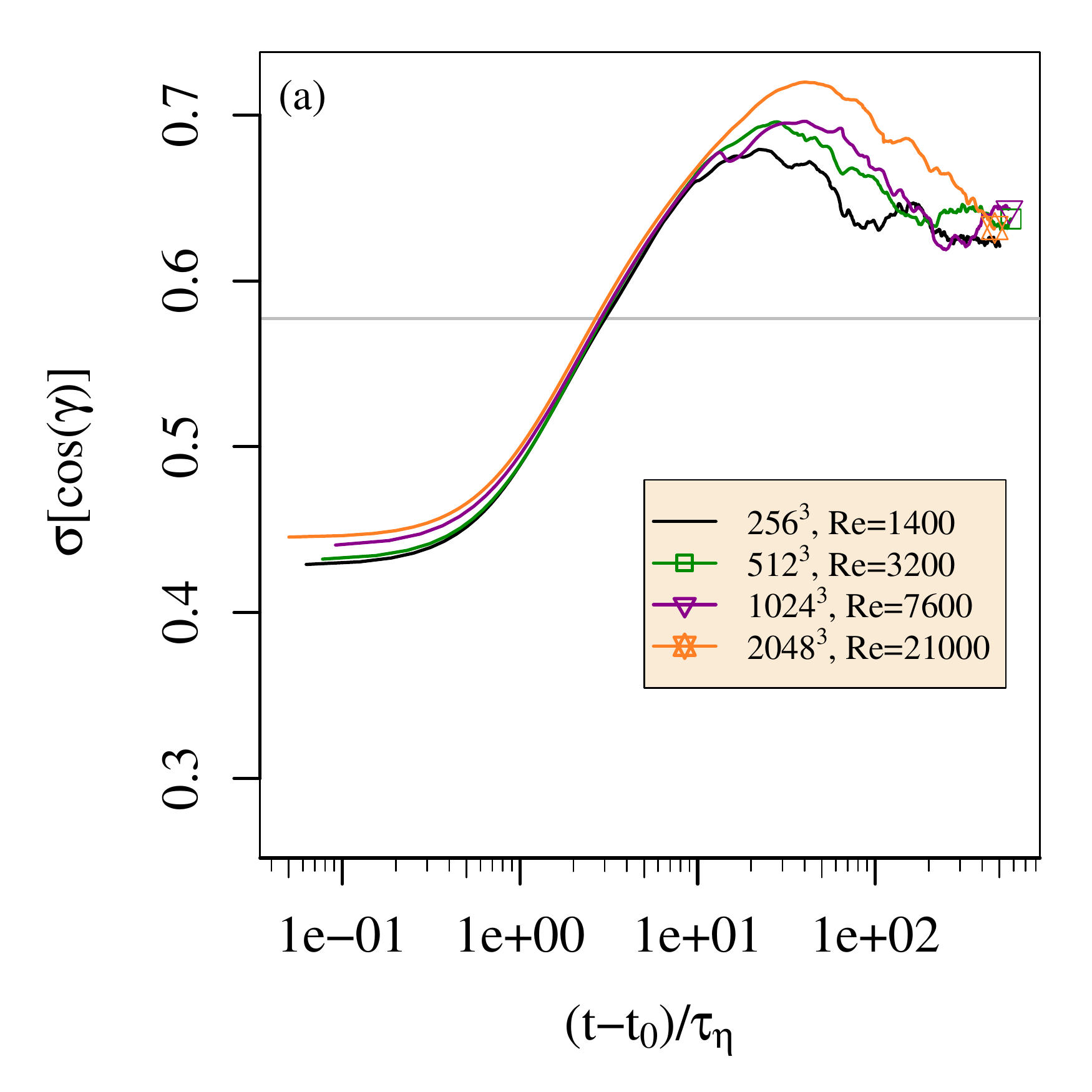}}
\resizebox{2.8in}{!}{\includegraphics{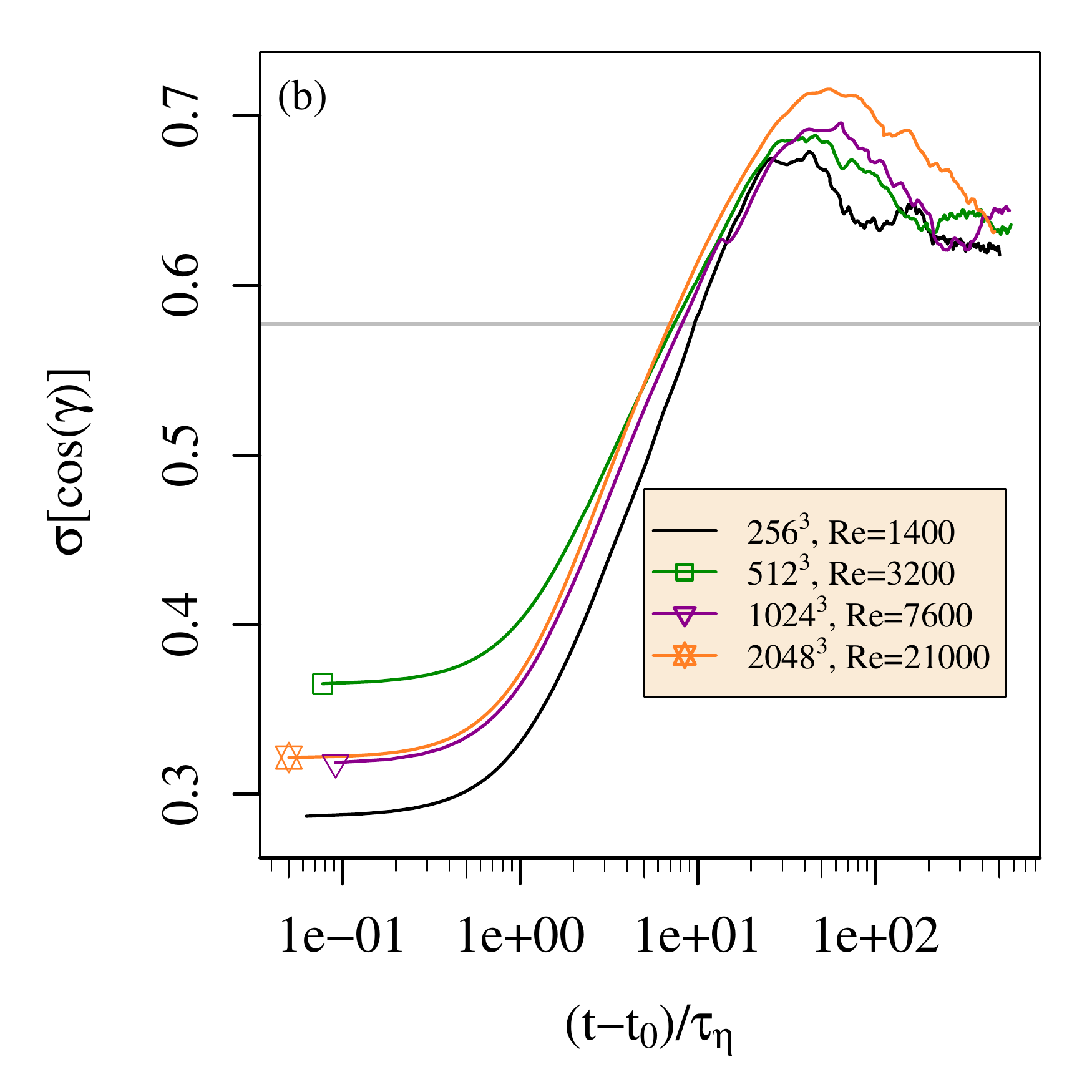}}
\caption{Standard deviation of the cosine of the magnetic alignment angle $\gamma$, for (a) perpendicular pairs,
and (b) parallel pairs.  These pairs of particles are initially
separated by 2 $\eta_{\mathsf{kol},\perp}$.  Time is given in units of the Kolmogorov time-scale $\mathsf{\tau_{\eta}}$.  The grey line indicates the value for an isotropic distribution of magnetic alignment angles.
\label{figxbrang}
}
\end{figure}

We find that higher order statistics, such as the standard deviation $\sigma[\cos{\gamma}]$ do exhibit clear trends for our simulations (see figure~\ref{figxbrang}).  If no magnetic alignment angles are preferred, then the distribution of $\cos{\gamma}$ is uniform on $[-1,1]$ and its standard deviation is $\sqrt{1/3}$, a point
  that is marked by a horizontal grey line in the figure.   For perpendicular pairs, at early times the distribution of magnetic alignment angles is concentrated around $\cos{\gamma}=0$ to a higher degree than for an isotropic case.   A rise in $\sigma[\cos{\gamma}]$ proceeds between roughly $2 \tau_{\eta}$ and $20  \tau_{\eta}$,  suggesting that as a pair of particles leaves the ballistic regime, the particles achieve a wider range of alignments with the local mean magnetic field, which also results in a lower alignment of pair separation with the relative velocity vector.
A more uniform distribution develops in this transient phase; the standard deviation passes through the point $\sqrt{1/3}$ at $3 \tau_{\eta}$ and continues to rise.   
  Near the beginning of the inertial subrange, the $\sigma[\cos{\gamma}]$ reaches a peak, at a state where the separation vector is preferentially aligned parallel or antiparallel to the local mean magnetic field.    For higher Reynolds number simulations, the peak in the standard deviation of $\cos{\gamma}$ is higher, corresponding to particle pairs that align more thoroughly with the local mean magnetic field.  Throughout the inertial subrange and diffusive regime, this alignment decreases, indicating that there is a progressive decorrelation; the value attained for long times is close, but does not reach,  the reference value of $\sqrt{1/3}$ for an isotropic distribution.
Parallel pairs are initialized so that their separation vectors are preferentially oriented to be anti-parallel to the mean magnetic field.  As this distribution becomes more uniform, $\sigma[\cos{\gamma}]$ grows. Eventually, the parallel pairs attain a mix of antiparallel and parallel alignments, and exhibit similar trends to the perpendicular pairs at longer times.

A comparison between the different alignment statistics in figures~\ref{figxbrang} and \ref{figcosvr} shows interesting similarities.
Parallel pairs reach the peak in $\langle \cos{\beta} \rangle $ slightly later than the perpendicular pairs, and the peak 
in $\sigma[\cos{\gamma}]$ also occurs at a slightly later time.   Although there is a lower level of alignment achieved for these parallel pairs, as measured by $\langle \cos{\beta} \rangle$, both types of pairs reach similar values of $\sigma[\cos{\gamma}]$. These observations point to two competing alignment processes
at work, stemming from the physical differences in the parallel and perpendicular directions.
Parallel pairs are initially strongly aligned with $B_0$, and they do not achieve a strong alignment between the separation vector and the relative velocity.
For perpendicular pairs, a strong alignment between the separation vector and relative velocity is readily developed at early times.  Then an increasing alignment with the local mean magnetic field reduces the alignment between the separation vector and relative velocity.  This ultimately slows the relative dispersion for both groups of particle pairs.

\section{Discussion and Conclusions \label{secdiscussion}}

We have produced Lagrangian statistics for incompressible homogeneous anisotropic MHD turbulence and examined how those statistics
change as a function of the Reynolds number.  Several of these diagnostics were investigated in earlier studies \citep{yeung2004relative,yeung2006reynolds,sawford2008reynolds} of incompressible homogeneous isotropic hydrodynamic turbulence at different Reynolds numbers.  From among the many diagnostics in these earlier works, we have selected those that are most significant to the physical
complexities of anisotropic MHD turbulence.    Additional diagnostics have also been evaluated to further clarify the Reynolds number dependence in this new physical setting. 

Anisotropic MHD turbulence is distinct from isotropic hydrodynamic turbulence in that the presence of a magnetic field interacts with the velocity of the flow, and a mean magnetic field introduces a global anisotropy in the dynamics.   
That anisotropy necessitates the use of a more restrictive resolution criterion, which we have introduced to assure that the energy spectra in the direction perpendicular to the magnetic field are resolved at the same level as the spectra in the direction aligned with the magnetic field.  Our examination of single-particle diffusion confirms that ballistic and diffusive scaling regimes apply in each of these physically distinct directions.  Two-particle dispersion also produces clear ballistic and diffusive regimes.  The diffusive regime exhibits a mildly super-diffusive scaling in both parallel and perpendicular directions.  During the inertial subrange, a scaling emerges as we examine larger Reynolds number simulations.  We quantify this scaling by examining the log derivatives of the dispersion curves.  In the perpendicular direction, for particles with separation $2 \eta_{\mathsf{kol},\perp}$ this scaling is close to 3; in the direction aligned with the magnetic field, this scaling is clearly larger than 3.  The skewness of the pair separation diagnostic reveals a higher level of intermittency with larger Reynolds number, and that intermittency is more pronounced in the direction aligned with the mean magnetic field.

The presence of a mean magnetic field also introduces large-scale Alfv\'enic fluctuations to the system, which interact with the turbulent dynamics.  We observe oscillations in the average square separation of particle pairs, clearly visible in the log-derivatives, which can be attributed to Alfv\'enic fluctuations.  Further work would be necessary to attempt to separate the effect of waves from the fundamental anisotropy.

To explore the anisotropy more deeply with a view toward theoretical modeling, we have examined the separation velocity of particle pairs.  The average separation velocity of particle pairs shows a dip following the ballistic
range and preceding the beginning of the inertial subrange, between $2 \tau_{\eta}$ and $10 \tau_{\eta}$.  This dip deepens for larger Reynolds number, and provides a transition
to the inertial subrange that is characteristically different from that of isotropic hydrodynamic turbulence.  
 We probe these results further by examining alignment statistics: the cosine of the alignment angle and the cosine of the magnetic alignment angle. 
Slightly preceding the dip, an increase in the standard deviation of the cosine of the magnetic alignment angle begins, and this increase continues throughout the dip period.  
 The dip period also correlates with a sharp drop in the average cosine of the alignment angle.  This indicates that there are two competing alignment processes at work, namely the alignment between the separation vector of a pair of particles and the relative velocity, and the alignment between the separation vector and local mean magnetic field.  The balance between these alignment processes is affected by the Reynolds number and by the anisotropy of the system.  These alignment statistics show clear trends with increasing Reynolds number.  A detailed evaluation of the strength of the mean magnetic field on alignment processes should be evaluated in future studies.

\backsection[Acknowledgements]{
This material is based upon work supported by the National Science Foundation under grant no. PHY-1907876.   A. Busse gratefully acknowledges support via a Leverhulme Trust Research Fellowship.  Simulations were performed on the Konrad and Gottfried computer systems
of the Norddeutsche Verbund zur F\"orderung des Hoch- und
H\"ochstleistungsrechnens (HLRN) by the project bep00051 ``Lagrangian studies of incompressible turbulence in plasmas''.  Part of this work was performed under the auspices of the U.S. Department of Energy by Lawrence Livermore National Laboratory under Contract DE-AC52-07NA27344.
}

\backsection[Author ORCID]{J. Pratt,  https://orcid.org/0000-0003-2707-3616; A. Busse,  https://orcid.org/0000-0002-3496-6036}

\backsection[Declaration of Interests]{
The authors report no conflict of interest.
}

\bibliographystyle{jfm}
\bibliography{postdoc}

\begin{thebibliography}{63}
\expandafter\ifx\csname natexlab\endcsname\relax\def\natexlab#1{#1}\fi
\def\au#1{#1} \def\ed#1{#1} \def\yr#1{#1}\def\at#1{#1}\def\jt#1{\textit{#1}}
  \def\bt#1{#1}\def\bvol#1{\textbf{#1}} \def\vol#1{#1} \def\pg#1{#1}
  \def\publ#1{#1}\def\arxiv#1{#1}\def\org#1{#1}\def\st#1{\textit{#1}}

\bibitem[Biferale {\em et~al.\/}(2005)Biferale, Boffetta, Celani, Devenish,
  Lanotte \& Toschi]{biferale2005lagrangian}
{\sc \au{Biferale, Luca}, \au{Boffetta, Guido}, \au{Celani, A}, \au{Devenish,
  BJ}, \au{Lanotte, A} \& \au{Toschi, Federico}} \yr{2005}  \at{Lagrangian
  statistics of particle pairs in homogeneous isotropic turbulence}.  \jt{Phys.
  Fluids}  \bvol{17}~(11),  \pg{115101}.

\bibitem[Biskamp(2003)]{biskampbook}
{\sc \au{Biskamp, D.}} \yr{2003} {\em Magnetohydrodynamic turbulence\/}.
  \publ{Cambridge University Press}.

\bibitem[Boffetta \& Sokolov(2002)]{boffetta2002statistics}
{\sc \au{Boffetta, G} \& \au{Sokolov, IM}} \yr{2002}  \at{Statistics of
  two-particle dispersion in two-dimensional turbulence}.  \jt{Phys. Fluids}
  \bvol{14}~(9),  \pg{3224--3232}.

\bibitem[Boldyrev(2005)]{boldyrev2005spectrum}
{\sc \au{Boldyrev, Stanislav}} \yr{2005}  \at{On the spectrum of
  magnetohydrodynamic turbulence}.  \jt{ApJ Lett.}  \bvol{626}~(1),  \pg{L37}.

\bibitem[Boldyrev(2006)]{boldymodelII}
{\sc \au{Boldyrev, S.}} \yr{2006}  \at{Spectrum of magnetohydrodynamic
  turbulence}.  \jt{Phys. Rev. Lett.}  \bvol{96},  \pg{115002}.

\bibitem[Boldyrev {\em et~al.\/}(2012)Boldyrev, Perez \&
  Zhdankin]{boldyrev2012residual}
{\sc \au{Boldyrev, Stanislav}, \au{Perez, Jean~Carlos} \& \au{Zhdankin,
  Vladimir}} \yr{2012} Residual energy in {MHD} turbulence and in the solar
  wind.  \bt{In {\em AIP Conference Proceedings\/}}, ,  \vol{vol. 1436},
  \pg{pp. 18--23}. American Institute of Physics.

\bibitem[Bourgoin(2015)]{bourgoin2015turbulent}
{\sc \au{Bourgoin, Micka{\"e}l}} \yr{2015}  \at{Turbulent pair dispersion as a
  ballistic cascade phenomenology}.  \jt{Journal of Fluid Mechanics}
  \bvol{772},  \pg{678--704}.

\bibitem[Brandenburg {\em et~al.\/}(2018)Brandenburg, Haugen, Li \&
  Subramanian]{brandenburg2018varying}
{\sc \au{Brandenburg, Axel}, \au{Haugen, Nils Erland~L}, \au{Li, Xiang-Yu} \&
  \au{Subramanian, K}} \yr{2018}  \at{Varying the forcing scale in low prandtl
  number dynamos}.  \jt{Monthly Notices of the Royal Astronomical Society}
  \bvol{479}~(2),  \pg{2827--2833}.

\bibitem[Busse \& M\"{u}ller(2008)]{busse_astnach}
{\sc \au{Busse, A.} \& \au{M\"{u}ller, W.-C.}} \yr{2008}  \at{Diffusion and
  dispersion in magnetohydrodynamic turbulence: The influence of mean magnetic
  fields}.  \jt{Astron. Nachr.}  \bvol{329}~(7),  \pg{714--416}.

\bibitem[Busse {\em et~al.\/}(2010)Busse, M{\"u}ller \&
  Gogoberidze]{busse2010lagrangian}
{\sc \au{Busse, Angela}, \au{M{\"u}ller, Wolf-Christian} \& \au{Gogoberidze,
  Grigol}} \yr{2010}  \at{Lagrangian frequency spectrum as a diagnostic for
  magnetohydrodynamic turbulence dynamics}.  \jt{Phys. Rev. Lett.}
  \bvol{105}~(23),  \pg{235005}.

\bibitem[Busse {\em et~al.\/}(2007)Busse, M\"{u}ller, Homann \&
  Grauer]{busse_hoho}
{\sc \au{Busse, Angela}, \au{M\"{u}ller, Wolf-Christian}, \au{Homann, Holger}
  \& \au{Grauer, Rainer}} \yr{2007}  \at{Statistics of passive tracers in
  three-dimensional magnetohydrodynamic turbulence}.  \jt{Phys. Plasmas}
  \bvol{14}~(12),  \pg{122303}.

\bibitem[Chandran(2008)]{chandran2008strong}
{\sc \au{Chandran, Benjamin~DG}} \yr{2008}  \at{Strong anisotropic {MHD}
  turbulence with cross helicity}.  \jt{Astrophys. J.}  \bvol{685}~(1),
  \pg{646}.

\bibitem[Chandran {\em et~al.\/}(2015)Chandran, Schekochihin \&
  Mallet]{chandranikintermitt}
{\sc \au{Chandran, B. D.~G.}, \au{Schekochihin, A.~A.} \& \au{Mallet, A.}}
  \yr{2015}  \at{Intermittency and alignment in strong {RMHD} turbulence}.
  \jt{ApJ}  \bvol{807}~(39),  \pg{16pp}.

\bibitem[Cho {\em et~al.\/}(2002)Cho, Lazarian \& Vishniac]{cho2002simulations}
{\sc \au{Cho, Jungyeon}, \au{Lazarian, Alex} \& \au{Vishniac, Ethan~T}}
  \yr{2002}  \at{Simulations of magnetohydrodynamic turbulence in a strongly
  magnetized medium}.  \jt{ApJ}  \bvol{564}~(1),  \pg{291}.

\bibitem[Cho \& Vishniac(2000)]{cho2000anisotropy}
{\sc \au{Cho, Jungyeon} \& \au{Vishniac, Ethan~T}} \yr{2000}  \at{The
  anisotropy of magnetohydrodynamic {A}lfv{\'e}nic turbulence}.  \jt{ApJ}
  \bvol{539}~(1),  \pg{273}.

\bibitem[Donzis {\em et~al.\/}(2008)Donzis, Yeung \&
  Sreenivasan]{donzis2008dissipation}
{\sc \au{Donzis, DA}, \au{Yeung, PK} \& \au{Sreenivasan, KR}} \yr{2008}
  \at{Dissipation and enstrophy in isotropic turbulence: resolution effects and
  scaling in direct numerical simulations}.  \jt{Phys. Fluids}  \bvol{20}~(4),
  \pg{045108}.

\bibitem[Dubbeldam {\em et~al.\/}(2009)Dubbeldam, Ford, Ellis \&
  Snurr]{dubbeldam2009new}
{\sc \au{Dubbeldam, David}, \au{Ford, Denise~C}, \au{Ellis, Donald~E} \&
  \au{Snurr, Randall~Q}} \yr{2009}  \at{A new perspective on the order-n
  algorithm for computing correlation functions}.  \jt{Mol. Simul.}
  \bvol{35}~(12-13),  \pg{1084--1097}.

\bibitem[Els{\"a}sser(1950)]{elsasser1950hydromagnetic}
{\sc \au{Els{\"a}sser, Walter~M}} \yr{1950}  \at{The hydromagnetic equations}.
  \jt{Physical Review}  \bvol{79}~(1),  \pg{183}.

\bibitem[Eyink {\em et~al.\/}(2013)Eyink, Vishniac, Lalescu, Aluie, Kanov,
  B{\"u}rger, Burns, Meneveau \& Szalay]{eyink2013flux}
{\sc \au{Eyink, Gregory}, \au{Vishniac, Ethan}, \au{Lalescu, Cristian},
  \au{Aluie, Hussein}, \au{Kanov, Kalin}, \au{B{\"u}rger, Kai}, \au{Burns,
  Randal}, \au{Meneveau, Charles} \& \au{Szalay, Alexander}} \yr{2013}
  \at{Flux-freezing breakdown in high-conductivity magnetohydrodynamic
  turbulence}.  \jt{Nature}  \bvol{497}~(7450),  \pg{466}.

\bibitem[Ghosal {\em et~al.\/}(1995)Ghosal, Lund, Moin \&
  Akselvoll]{ghosal1995dynamic}
{\sc \au{Ghosal, Sandip}, \au{Lund, Thomas~S}, \au{Moin, Parviz} \&
  \au{Akselvoll, K}} \yr{1995}  \at{A dynamic localization model for large-eddy
  simulation of turbulent flows}.  \jt{J. Fluid Mech}  \bvol{286},
  \pg{229--255}.

\bibitem[Goldreich \& Sridhar(1995{\natexlab{{\em a\/}}})]{goldreich1995toward}
{\sc \au{Goldreich, P} \& \au{Sridhar, S}} \yr{1995{\natexlab{{\em a\/}}}}
  \at{Toward a theory of interstellar turbulence. 2: {S}trong {A}lfv\'enic
  turbulence}.  \jt{ApJ}  \bvol{438},  \pg{763--775}.

\bibitem[Goldreich \& Sridhar(1995{\natexlab{{\em b\/}}})]{goldreich_sridhar2}
{\sc \au{Goldreich, P.} \& \au{Sridhar, S.}} \yr{1995{\natexlab{{\em b\/}}}}
  \at{Toward a theory of interstellar turbulence. {II.\ Strong A}lfv\'enic
  turbulence}.  \jt{Astrophysical Journal}  \bvol{438},  \pg{763--775}.

\bibitem[Heiles \& Troland(2005)]{heiles2005millennium}
{\sc \au{Heiles, Carl} \& \au{Troland, TH}} \yr{2005}  \at{The millennium
  arecibo 21 centimeter absorption-line survey. iv. statistics of magnetic
  field, column density, and turbulence}.  \jt{ApJ}  \bvol{624}~(2),  \pg{773}.

\bibitem[Hennebelle \& Falgarone(2012)]{hennebelle2012turbulent}
{\sc \au{Hennebelle, Patrick} \& \au{Falgarone, Edith}} \yr{2012}
  \at{Turbulent molecular clouds}.  \jt{The Astronomy and Astrophysics Review}
  \bvol{20}~(1),  \pg{55}.

\bibitem[Heyer \& Brunt(2004)]{heyer2004universality}
{\sc \au{Heyer, Mark~H} \& \au{Brunt, Christopher~M}} \yr{2004}  \at{The
  universality of turbulence in galactic molecular clouds}.  \jt{ApJ Lett.}
  \bvol{615}~(1),  \pg{L45}.

\bibitem[Homann {\em et~al.\/}(2007{\natexlab{{\em a\/}}})Homann, Dreher \&
  Grauer]{homann2007impact}
{\sc \au{Homann, Holger}, \au{Dreher, J{\"u}rgen} \& \au{Grauer, Rainer}}
  \yr{2007{\natexlab{{\em a\/}}}}  \at{Impact of the floating-point precision
  and interpolation scheme on the results of dns of turbulence by
  pseudo-spectral codes}.  \jt{Comput. Phys. Commun.}  \bvol{177}~(7),
  \pg{560--565}.

\bibitem[Homann {\em et~al.\/}(2007{\natexlab{{\em b\/}}})Homann, Grauer, Busse
  \& M{\"u}ller]{homann2007lagrangian}
{\sc \au{Homann, Holger}, \au{Grauer, Rainer}, \au{Busse, Angela} \&
  \au{M{\"u}ller, Wolf-Christian}} \yr{2007{\natexlab{{\em b\/}}}}
  \at{Lagrangian statistics of {N}avier--{S}tokes and {MHD} turbulence}.
  \jt{Journal of Plasma Physics}  \bvol{73}~(6),  \pg{821--830}.

\bibitem[Howes(2015)]{howes2015inherently}
{\sc \au{Howes, Gregory~G}} \yr{2015}  \at{The inherently three-dimensional
  nature of magnetized plasma turbulence}.  \jt{J Plasma Phys.}  \bvol{81}~(2).

\bibitem[Ishihara {\em et~al.\/}(2009)Ishihara, Gotoh \&
  Kaneda]{ishihara_gotoh_kaneda}
{\sc \au{Ishihara, T.}, \au{Gotoh, T.} \& \au{Kaneda, Y.}} \yr{2009}  \at{Study
  of high-{R}eynolds-number isotropic turbulence by direct numerical
  simulation}.  \jt{Annu. Rev. Fluid Mech.}  \bvol{41},  \pg{165--180}.

\bibitem[Lekien \& Marsden(2005)]{lekien2005tricubic}
{\sc \au{Lekien, Francois} \& \au{Marsden, J}} \yr{2005}  \at{Tricubic
  interpolation in three dimensions}.  \jt{International Journal for Numerical
  Methods in Engineering}  \bvol{63}~(3),  \pg{455--471}.

\bibitem[Malik \& Hussain(2021)]{malik2021new}
{\sc \au{Malik, Nadeem~A} \& \au{Hussain, Fazle}} \yr{2021}  \at{New scaling
  laws predicting turbulent particle pair diffusion, overcoming the limitations
  of the prevalent {R}ichardson--{O}bukhov theory}.  \jt{Phys. Fluids}
  \bvol{33}~(3),  \pg{035135}.

\bibitem[Mallet \& Schekochihin(2017)]{malletstrongMHDmod}
{\sc \au{Mallet, A.} \& \au{Schekochihin, A.~A.}} \yr{2017}  \at{A statistical
  model of three-dimensional anisotropy and intermittency in strong
  {A}lfv\'enic turbulence}.  \jt{Monthly Notices of the Royal Astronomical
  Society}  \bvol{466},  \pg{3918--3927}.

\bibitem[Mallet {\em et~al.\/}(2015)Mallet, Schekochihin \&
  Chandran]{malletrefinedcb}
{\sc \au{Mallet, A.}, \au{Schekochihin, A.~A.} \& \au{Chandran, B. D.~G.}}
  \yr{2015}  \at{Refined critical balance in strong {A}lfv\'enic turbulence}.
  \jt{Monthly Notices of the Royal Astronomical Society}  \bvol{449},
  \pg{L77--L81}.

\bibitem[McKay {\em et~al.\/}(2017)McKay, Linkmann, Clark, Chalupa \&
  Berera]{mckay2017comparison}
{\sc \au{McKay, Mairi~E}, \au{Linkmann, Moritz}, \au{Clark, Daniel},
  \au{Chalupa, Adam~A} \& \au{Berera, Arjun}} \yr{2017}  \at{Comparison of
  forcing functions in magnetohydrodynamics}.  \jt{Physical Review Fluids}
  \bvol{2}~(11),  \pg{114604}.

\bibitem[M{\"u}ller \& Busse(2007)]{muller2007diffusion}
{\sc \au{M{\"u}ller, W-C} \& \au{Busse, Angela}} \yr{2007}  \at{Diffusion and
  dispersion of passive tracers: {N}avier-{S}tokes vs. {MHD} turbulence}.
  \jt{EPL (Europhysics Letters)}  \bvol{78}~(1),  \pg{14003}.

\bibitem[M{\"u}ller \& Grappin(2005)]{muller2005spectral}
{\sc \au{M{\"u}ller, Wolf-Christian} \& \au{Grappin, Roland}} \yr{2005}
  \at{Spectral energy dynamics in magnetohydrodynamic turbulence}.  \jt{Phys.
  Rev. Lett.}  \bvol{95}~(11),  \pg{114502}.

\bibitem[M{\"u}ller {\em et~al.\/}(2012)M{\"u}ller, Malapaka \&
  Busse]{muller2012inverse}
{\sc \au{M{\"u}ller, Wolf-Christian}, \au{Malapaka, Shiva~Kumar} \& \au{Busse,
  Angela}} \yr{2012}  \at{Inverse cascade of magnetic helicity in
  magnetohydrodynamic turbulence}.  \jt{Phys Rev E}  \bvol{85}~(1),
  \pg{015302}.

\bibitem[Perez \& Boldyrev(2007)]{perez2007weak}
{\sc \au{Perez, Jean~Carlos} \& \au{Boldyrev, Stanislav}} \yr{2007}  \at{On
  weak and strong magnetohydrodynamic turbulence}.  \jt{ApJ Lett.}
  \bvol{672}~(1),  \pg{L61}.

\bibitem[Perez {\em et~al.\/}(2014)Perez, Mason, Boldyrev \&
  Cattaneo]{perez2014scaling}
{\sc \au{Perez, Jean~Carlos}, \au{Mason, Joanne}, \au{Boldyrev, Stanislav} \&
  \au{Cattaneo, Fausto}} \yr{2014}  \at{Scaling properties of small-scale
  fluctuations in magnetohydrodynamic turbulence}.  \jt{ApJ Lett}
  \bvol{793}~(1),  \pg{L13}.

\bibitem[Pope(2000)]{pope2000turbulent}
{\sc \au{Pope, Stephen~B}} \yr{2000} {\em Turbulent flows\/}.  \publ{Cambridge
  University Press}.

\bibitem[Pratt {\em et~al.\/}(2017)Pratt, Busse, M{\"u}ller, Watkins \&
  Chapman]{pratt2017extreme}
{\sc \au{Pratt, J}, \au{Busse, A}, \au{M{\"u}ller, WC}, \au{Watkins, NW} \&
  \au{Chapman, Sandra~C}} \yr{2017}  \at{Extreme-value statistics from
  {L}agrangian convex hull analysis for homogeneous turbulent {B}oussinesq
  convection and {MHD} convection}.  \jt{New Journal of Physics}
  \bvol{19}~(6),  \pg{065006}.

\bibitem[Pratt {\em et~al.\/}(2020{\natexlab{{\em a\/}}})Pratt, Busse \&
  M{\"u}ller]{pratt2020intermittency}
{\sc \au{Pratt, J}, \au{Busse, A} \& \au{M{\"u}ller, W-C}}
  \yr{2020{\natexlab{{\em a\/}}}} Intermittency of many-particle dispersion in
  anisotropic magnetohydrodynamic turbulence.  \bt{In {\em Journal of Physics:
  Conference Series\/}}, ,  \vol{vol. 1620 (1)},  \pg{p. 012015}. IOP
  Publishing.

\bibitem[Pratt {\em et~al.\/}(2020{\natexlab{{\em b\/}}})Pratt, Busse \&
  M{\"u}ller]{pratt2020lagrangian}
{\sc \au{Pratt, Jane}, \au{Busse, Angela} \& \au{M{\"u}ller, W-C}}
  \yr{2020{\natexlab{{\em b\/}}}}  \at{Lagrangian statistics for dispersion in
  magnetohydrodynamic turbulence}.  \jt{Journal of Geophysical Research: Space
  Physics}  \bvol{125}~(11),  \pg{e2020JA028245}.

\bibitem[Richardson(1926)]{richardson1926atmospheric}
{\sc \au{Richardson, Lewis~Fry}} \yr{1926}  \at{Atmospheric diffusion shown on
  a distance-neighbour graph}.  \jt{Proceedings of the Royal Society of London.
  Series A, Containing Papers of a Mathematical and Physical Character}
  \bvol{110}~(756),  \pg{709--737}.

\bibitem[Salazar \& Collins(2009)]{salazar2009two}
{\sc \au{Salazar, Juan~PLC} \& \au{Collins, Lance~R}} \yr{2009}
  \at{Two-particle dispersion in isotropic turbulent flows}.  \jt{Annu. Rev.
  Fluid Mech.}  \bvol{41},  \pg{405--432}.

\bibitem[Sawford(1991)]{sawford1991reynolds}
{\sc \au{Sawford, BL}} \yr{1991}  \at{Reynolds number effects in {L}agrangian
  stochastic models of turbulent dispersion}.  \jt{Phys. Fluids A: Fluid
  Dynamics}  \bvol{3}~(6),  \pg{1577--1586}.

\bibitem[Sawford(2001)]{sawford2001turbulent}
{\sc \au{Sawford, Brian}} \yr{2001}  \at{Turbulent relative dispersion}.
  \jt{Annu. Rev. Fluid Mech.}  \bvol{33}~(1),  \pg{289--317}.

\bibitem[Sawford {\em et~al.\/}(2008)Sawford, Yeung \&
  Hackl]{sawford2008reynolds}
{\sc \au{Sawford, Brian~L}, \au{Yeung, PK} \& \au{Hackl, Jason~F}} \yr{2008}
  \at{Reynolds number dependence of relative dispersion statistics in isotropic
  turbulence}.  \jt{Phys. Fluids}  \bvol{20}~(6),  \pg{065111}.

\bibitem[Schekochihin {\em et~al.\/}(2008)Schekochihin, Cowley \&
  Yousef]{schekochihin2008mhd}
{\sc \au{Schekochihin, Alexander~A}, \au{Cowley, Steven~C} \& \au{Yousef,
  Tarek~A}} \yr{2008} {MHD} turbulence: Nonlocal, anisotropic, nonuniversal?
  \bt{In {\em IUTAM Symposium on computational physics and new perspectives in
  turbulence\/}},  \pg{pp. 347--354}. Springer.

\bibitem[Sokolov(1999)]{sokolov1999two}
{\sc \au{Sokolov, IM}} \yr{1999}  \at{Two-particle dispersion by correlated
  random velocity fields}.  \jt{Phys. Rev. E}  \bvol{60}~(5),  \pg{5528}.

\bibitem[Taylor(1922)]{taylor1922diffusion}
{\sc \au{Taylor, Geoffrey~I}} \yr{1922}  \at{Diffusion by continuous
  movements}.  \jt{Proceedings of the {L}ondon {M}athematical {S}ociety}
  \bvol{2}~(1),  \pg{196--212}.

\bibitem[Utomo {\em et~al.\/}(2019)Utomo, Blitz \& Falgarone]{utomo2019origin}
{\sc \au{Utomo, Dyas}, \au{Blitz, Leo} \& \au{Falgarone, Edith}} \yr{2019}
  \at{The origin of interstellar turbulence in {M33}}.  \jt{ApJ}
  \bvol{871}~(1),  \pg{17}.

\bibitem[Verdini \& Grappin(2012)]{verdini2012transition}
{\sc \au{Verdini, Andrea} \& \au{Grappin, Roland}} \yr{2012}  \at{Transition
  from weak to strong cascade in {MHD} turbulence}.  \jt{Phys. Rev. Lett.}
  \bvol{109}~(2),  \pg{025004}.

\bibitem[Verdini {\em et~al.\/}(2015)Verdini, Grappin, Hellinger, Landi \&
  M{\"u}ller]{verdini2015anisotropy}
{\sc \au{Verdini, Andrea}, \au{Grappin, Roland}, \au{Hellinger, Petr},
  \au{Landi, Simone} \& \au{M{\"u}ller, Wolf~Christian}} \yr{2015}
  \at{Anisotropy of third-order structure functions in {MHD} turbulence}.
  \jt{ApJ}  \bvol{804}~(2),  \pg{119}.

\bibitem[Vorobev {\em et~al.\/}(2005)Vorobev, Zikanov, Davidson \&
  Knaepen]{vorobev2005anisotropy}
{\sc \au{Vorobev, Anatoliy}, \au{Zikanov, Oleg}, \au{Davidson, Peter~A} \&
  \au{Knaepen, Bernard}} \yr{2005}  \at{Anisotropy of magnetohydrodynamic
  turbulence at low magnetic {R}eynolds number}.  \jt{Phys. Fluids}
  \bvol{17}~(12),  \pg{125105}.

\bibitem[Williamson(1980)]{will80}
{\sc \au{Williamson, J.~H.}} \yr{1980}  \at{Low-storage {R}unge-{K}utta
  schemes}.  \jt{J. Comput. Phys.}  \bvol{35}~(1),  \pg{48 -- 56}.

\bibitem[Yeung(1994)]{yeung1994direct}
{\sc \au{Yeung, PK}} \yr{1994}  \at{Direct numerical simulation of two-particle
  relative diffusion in isotropic turbulence}.  \jt{Phys. Fluids}
  \bvol{6}~(10),  \pg{3416--3428}.

\bibitem[Yeung \& Borgas(2004)]{yeung2004relative}
{\sc \au{Yeung, PK} \& \au{Borgas, Michael~S}} \yr{2004}  \at{Relative
  dispersion in isotropic turbulence. part 1. direct numerical simulations and
  {R}eynolds-number dependence}.  \jt{Journal of Fluid Mechanics}  \bvol{503},
  \pg{93--124}.

\bibitem[Yeung \& Pope(1989)]{yeung1989lagrangian}
{\sc \au{Yeung, PK} \& \au{Pope, SB}} \yr{1989}  \at{Lagrangian statistics from
  direct numerical simulations of isotropic turbulence}.  \jt{J. Fluid Mech.}
  \bvol{207},  \pg{531--586}.

\bibitem[Yeung {\em et~al.\/}(2006)Yeung, Pope \& Sawford]{yeung2006reynolds}
{\sc \au{Yeung, PK}, \au{Pope, Stephen~B} \& \au{Sawford, Brian~Lewis}}
  \yr{2006}  \at{Reynolds number dependence of lagrangian statistics in large
  numerical simulations of isotropic turbulence}.  \jt{Journal of Turbulence}
  ~(7),  \pg{N58}.

\bibitem[Yeung {\em et~al.\/}(2018)Yeung, Sreenivasan \&
  Pope]{yeung2018effects}
{\sc \au{Yeung, PK}, \au{Sreenivasan, KR} \& \au{Pope, SB}} \yr{2018}
  \at{Effects of finite spatial and temporal resolution in direct numerical
  simulations of incompressible isotropic turbulence}.  \jt{Physical Review
  Fluids}  \bvol{3}~(6),  \pg{064603}.

\bibitem[Zahn(1993)]{zahn1993mixing}
{\sc \au{Zahn, J-P}} \yr{1993}  \at{Mixing processes and stellar evolution}.
  \jt{Space Science Reviews}  \bvol{66}~(1-4),  \pg{285--297}.

\bibitem[Zimbardo {\em et~al.\/}(2010)Zimbardo, Greco, Sorriso-Valvo, Perri,
  V{\"o}r{\"o}s, Aburjania, Chargazia \& Alexandrova]{zimbardo2010magnetic}
{\sc \au{Zimbardo, G}, \au{Greco, A}, \au{Sorriso-Valvo, L}, \au{Perri, S},
  \au{V{\"o}r{\"o}s, Z}, \au{Aburjania, G}, \au{Chargazia, K} \&
  \au{Alexandrova, O}} \yr{2010}  \at{Magnetic turbulence in the geospace
  environment}.  \jt{Space Science Reviews}  \bvol{156}~(1-4),  \pg{89--134}.

\end{thebibliography}

\end{document}